\documentclass[10pt,a4paper,twocolumn,english,pra,aps,showpacs,floatfix,groupedaddress,superscriptaddress]{revtex4-1}

\usepackage{graphicx}
\usepackage{epsfig}
\usepackage[english]{babel}
\usepackage{amsmath}
\usepackage{amssymb}
\usepackage{amsfonts}
\usepackage{longtable}
\usepackage{dcolumn}% Align table columns on decimal point
\usepackage{bm}
\usepackage{bbm}
\usepackage{nicefrac}
\usepackage{color,array}
\usepackage{colortbl}

\usepackage{dsfont}

\begin{document}
\title{Efficient algorithms for the dynamics of large and 
infinite classical central spin models}

\author{Benedikt Fauseweh}
\email{benedikt.fauseweh@tu-dortmund.de}
\affiliation{Lehrstuhl f\"{u}r Theoretische Physik I, Technische Universit\"at Dortmund, Otto-Hahn Stra\ss{}e 4, 44221 Dortmund, Germany}

\author{Philipp Schering}
\affiliation{Lehrstuhl f\"{u}r Theoretische Physik I, Technische Universit\"at Dortmund, Otto-Hahn Stra\ss{}e 4, 44221 Dortmund, Germany}

\author{Jan H\"udepohl}
\affiliation{Lehrstuhl f\"{u}r Theoretische Physik I, Technische Universit\"at Dortmund, Otto-Hahn Stra\ss{}e 4, 44221 Dortmund, Germany}

\author{G\"otz S.\ Uhrig}
\email{goetz.uhrig@tu-dortmund.de}
\affiliation{Lehrstuhl f\"{u}r Theoretische Physik I, Technische Universit\"at Dortmund, Otto-Hahn Stra\ss{}e 4, 44221 Dortmund, Germany}

\date{\rm\today}

\begin{abstract}
We investigate the time dependence of correlation functions in the central spin
 model, which describes the electron or hole spin confined in a quantum dot, interacting with a bath of nuclear spins forming the Overhauser field. For large baths, a classical description of the model yields quantitatively correct results. We develop and apply various algorithms in order to capture the long-time  limit of the central spin for bath sizes from 1000 to infinitely many bath spins.  Representing the Overhauser field in terms of orthogonal polynomials, we show that
a carefully reduced set of differential equations is sufficient to compute the spin correlations of the full problem up to very long times, for instance up to $10^5\hbar/J_\mathrm{Q}$ where $J_\mathrm{Q}$ is the natural energy unit of the system. This technical progress renders an analysis of the model with experimentally relevant parameters possible. We benchmark the results of the algorithms with exact data for a small number of bath spins and we predict how the long-time
correlations behave for different effective numbers of bath spins.
\end{abstract}

%% PACSes
% 03.65.Yz Decoherence; open systems; quantum statistical methods 
% 78.67.Hc Quantum dots 
% 72.25.Rb Spin relaxation and scattering 
% 03.65.Sq Semiclassical theories and applications 

\maketitle

\section{Introduction}
\label{sec.intro}

A localized spin with $S=1/2$ is the simplest realization of a small
quantum system. At the same time, it can serve as a quantum bit in the 
context of quantum information processing. \cite{niels00}. 
There are many experimental realizations,
for instance by impurities in solid state systems \cite{jelez06}.
Due to the many possibilities to design semiconductor nanostructures
a particularly interesting realization of a spin $S=1/2$ system
is the spin  of an excess electron or hole in single quantum dots
\cite{loss98,hanso07,urbas13} or ensembles of quantum dots 
\cite{schli03,greil06b,greil07a}.

In quantum dots, the dominating coupling of the electronic $S=1/2$ is via
its hyperfine coupling to the nuclear spins, which are almost omnipresent
in the generically used semiconductors \cite{schli03,merku02}.
The ensemble of nuclear spins acts as a bath on the electronic spin.
A suitable model to describe the dynamics of the electronic spin
is the central spin model introduced by Gaudin as a case of 
a correlated model solvable by means of the Bethe ansatz 
\cite{gaudi76,gaudi83,bortz07b,bortz10b,farib13a}.

In spite of the analytic solution, the complex dynamics
in the central spin model poses a challenging issue even today. 
The exact solution is only tractable for fairly small systems of
about 30-40 spins while experimental quantum dots host
about $10^4$-$10^5$ nuclear spins within the localization volume
of the electronic spin \cite{merku02,schli03,lee05,petro08};
this number will be called the effective number of spins $N_\text{eff}$.
The total number of nuclear spins, which couple to the central spin,
however weakly, is even much larger and can safely be regarded as infinity.

In view of the large systems and the long times (up to minutes) to be studied
many complementary theoretical techniques have been employed. 
Besides the already mentioned Bethe ansatz, 
exact diagonalization \cite{schli03,cywin10}, Chebyshev expansion (CE)
\cite{dobro03a,dobro03b,hackm14a}, or a direct evolution
of the density matrices via the Liouvillean \cite{beuge16} can be used for small 
systems of about 20 spins, but up to long times. Density-matrix renormalization group (DMRG)
can tackle much larger systems up to about 1000 spins, but is
restricted to short times up to about $40\hbar/J_\mathrm{Q}$ \cite{stane13,stane14b,grave16}.
The strict limit of infinite times, i.e., of persisting correlations
has been tackled by mathematically rigorous bounds \cite{uhrig14a,seife16}.
Techniques based on non-Markovian master equations give access to
large bath sizes, but are well justified only for sufficiently strong
external fields 
\cite{khaet02,khaet03,coish04,breue04,fisch07,ferra08,coish10,barne12}.
The same holds for approaches based on equations of motion \cite{deng06,deng08}. 
Cluster expansion techniques represent another
powerful approach restricted by the maximum cluster size kept,
which translates into a certain time threshold up to which the results
are reliable \cite{witze05,witze06,maze08,yang08a,yang09a,witze12}.

Real-time dynamics has been frequently studied in semiclassical or classical models. One approach is to replace the bath by an effective
time-dependent field \cite{erlin02,erlin04,merku02,stane13}.
As a first approximation, the bath may be regarded as frozen, i.e., 
the Overhauser field is constant.  Subsequently, random fluctuations of the bath
due to the interaction with the central spin can be included \cite{merku02}.
Assuming that the Overhauser field can be described as a stochastic field, the fluctuations of the central spin can be found from solution of the Bloch equation of the Langevin type \cite{glazo12a,hackm14a}.

Furthermore, it was argued that the saddle-point approximation of the spin-coherent path integral representation describes the central spin dynamics well because the quantum fluctuations become less important
for large numbers of bath spins \cite{chen07}. Similarly, the
so-called $P$ representation of the density matrix with time-dependent
mean-field theory amounts to solving essentially classical
equations of motion \cite{alhas06,zhang06}. Previously, the comparison
of DMRG and CE data with Gaussian weighted classical simulations 
showed very good agreement \cite{stane14b}. This approach is backed 
by the analytical argument that the Overhauser field stemming 
from a very large number of quantum spins behaves 
like a classical variable \cite{stane14b}. 

Still, even the simulation of the classical model represents 
an impossible task for $10^5$ spins. It is the purpose of 
the present paper to establish efficient algorithms, which
enable us to meet this challenge successfully. 
Thus we can now explore time scales for large bath sizes, which previously were inaccessible.  In this way, we establish that the long-time 
behavior of the system is governed by a low-energy scale different from the 
energy scale $J_\text{Q}$. This low-energy scale is proportional to the inverse 
number of effectively coupled bath spins.

The paper is set up as follows. First, the model is introduced
in all its details in Sec.\ \ref{sec.model}. In Sec.\ \ref{sec.expansion},
three approaches to the classical simulation are introduced, of which
two work very well. The results are shown and compared in
Sec.\ \ref{sec.results}. A particular focus lies on the long-time
behavior and its scaling with the number of bath spins.
Finally, the conclusions are drawn in Sec.\ \ref{sec.conclusions}.

\section{Model}
\label{sec.model}

The Hamiltonian operator of the central spin model is given by
\begin{align}
H = \vec{\hat{S}}_0 \cdot \sum\limits_{i=1}^{N} J_i \vec{\hat{S}}_i , 
\end{align}
where $S_0$ is the central $S=1/2$ spin, which is coupled in a star configuration to 
$N$ $S=1/2$ spins, which form the bath.
We remind the reader that in a realistic quantum dot the nuclear spins
do not have $S=1/2$. However, for simplicity we consider this case here.
As we will show shortly, the classical treatment only requires very limited
information about the bath spins anyway
so that this restriction is not harmful.

The hyperfine interaction between the central spin and the spin $i$ is assumed to be isotropic and given by $J_i$. The bath spins act via $J_i$ as an effective magnetic field on the central spin. This field resulting from all bath spins is called the Overhauser field and is denoted by
\begin{align}
\vec{\hat{B}} =  \sum\limits_{i=1}^{N} J_i \vec{\hat{S}}_i .
\end{align}
In a quantum dot, the central spin represents the single electron (or hole) spin interacting with a bath of nuclear spins. Since the dipole-dipole interaction between the nuclear spins is negligibly small compared to the hyperfine 
interaction with the electron spin, it is not considered \cite{merku02,schli03}.

The hyperfine interaction is proportional to the probability density of the electronic wave function at the location of the nuclear spin. 
For a Gaussian wave function in two dimensions this leads to an 
an exponential distribution of the exchange interaction (see Appendix \
\ref{app.weight} for details)
\begin{align}
\label{eq:exponen}
J_i = C \exp(-i \gamma)  ,
\end{align}
where $C$ is an energy constant and the index $i$ runs from 1 to $N$.
Note that $N$ can be set to infinity. Similar distributions have been studied before \cite{farib13b,seife16}. 
In our concrete calculations we use the energy $J_\mathrm{Q}$ defined by
\begin{align}
\label{eq:jnorm}
J_\mathrm{Q}^2 :=\sum\limits_{i=1}^N J_i^2
\end{align}
as the natural energy unit, i.e., we determine $C$ so that $J_\mathrm{Q}=1$ holds.

What is the significance of the parameter $\gamma$? Besides $J_\mathrm{Q}$ 
we introduce the sum of all couplings 
\begin{align}
J_\mathrm{S} :=\sum\limits_{i=1}^N J_i . 
\end{align}
to clarify this question. Then we consider the simplest distribution for comparison, namely a uniform one where all $J_i=C$ implying $J_\mathrm{S}=CN$
and $J_\mathrm{Q}^2=C^2N$ so that the ratio $J_\mathrm{S}^2/J_\mathrm{Q}^2=N$ yields the number
of spins. For the distribution \eqref{eq:exponen} one has 
\begin{subequations}
\label{eq:SQ}
\begin{align}
J_\mathrm{S} &=\sum\limits_{i=1}^\infty C\exp(-i\gamma) \\
&= C\exp(-\gamma)/[1-\exp(-\gamma)]
\\
J_\mathrm{Q}^2 &=\sum\limits_{i=1}^\infty C^2\exp(-i2\gamma) \\
&= C^2\exp(-2\gamma)/[1-\exp(-2\gamma)]
\end{align}
\end{subequations}
for the infinite bath.
Note that for large baths for which $N\gamma\gg 1$ holds there
is only an exponentially small difference between large finite $N$
and $N=\infty$. From \eqref{eq:SQ} we deduce
\begin{subequations}
\begin{align}
N_\mathrm{eff} &:=  \frac{J_\mathrm{S}^2}{J_\mathrm{Q}^2} 
\\
&= \frac{1-\exp(-2\gamma)}{[1-\exp(-\gamma)]^2} 
\\
&=\frac{2}{\gamma} + \mathcal{O}(\gamma^0),
\end{align}
\end{subequations}
where the last relation holds for small values of $\gamma$. We deduce
that the \emph{effective} number of spins is not infinity even if
$N=\infty$ holds, but proportional to the inverse of $\gamma=2/N_\mathrm{eff}$.
This implies that $\gamma\approx 10^{-5}$ for generic quantum dots 
\cite{merku02,lee05,petro08}.
In contrast, for large $\gamma$, the dynamics of the central spin is determined by 
a small number of bath spins and can be determined using a fully quantum 
mechanical description \cite{beuge16}.

In this paper, we are interested in the 
autocorrelation function of the central spin
\begin{align}
S(t):=\left\langle \hat{S}_0^z(t) \hat{S}_0^z(0) \right\rangle
\end{align}
for small values of $\gamma$. Note that the correlation $S(t)$ is fully equivalent 
to the time evolution of the expectation values of
$\frac{1}{2}\hat{S}_0^z(t)$ evaluated for an initial $\uparrow$
central spin with $S_0=1/2$ as we
study here. We focus on the case where the bath is initially completely disordered,
which corresponds to infinite temperature or equivalently to the fact 
that the density matrix is proportional to the identity
\begin{align}
\hat{\rho} = \frac{1}{Z}  \mathds{\hat{1}},
\end{align}
where $Z$ is the dimension of the total Hilbert space normalizing
the density matrix.
This is a realistic experimental scenario because the characteristic
thermal energy $k_\mathrm{B}T$ is generically at least one order of magnitude larger than the internal energy scale $J_\mathrm{Q}$ of the quantum dot.

For this case, it was shown in Ref.\ \cite{stane14b} that the quantum mechanical expectation value can be computed reliably by the mean values of a simulation of classical vectors starting from Gaussian random fields for each component of the bath spin vectors and of the central spin.
 Thus, we replace the quantum mechanical operators $\vec{\hat{S}}_i$ by real-valued time-dependent vectors $\vec{S}_i(t)$. The equations of motion read
\begin{align}
\label{eq:exact_central_spin}
\frac{\mathrm{d}}{\mathrm{d}t} \vec{S}_0 = \vec{B} \times \vec{S}_0
\end{align}
for the central spin and
\begin{align}
\label{eq:exact_bath_spin}
\frac{\mathrm{d}}{\mathrm{d}t} \vec{S}_i = J_i \vec{S}_0 \times \vec{S}_i
\end{align}
for the bath spins.
To simulate the quantum mechanical expectation values, the initial condition for the components of the vectors are drawn randomly with average value $\mu=0$ and variance $\sigma^2$ determined such that it coincides with the quantum mechanical
expectation value
\begin{align}
\left\langle \hat{S}_i^\alpha(0) \hat{S}_j^\beta(0) \right\rangle  =
 \delta_{\alpha\beta} \delta_{ij} \frac{1}{4} \ , \quad \alpha, \beta \in 
\left\lbrace x,y,z \right\rbrace .
\end{align}
Thus, for a single spin component the variance $\sigma^2$ is given by $1/4$.

In practice one must average over an appropriate large number of configurations
of the random fields. About $10^6$
simulations are enough to reduce the relative statistical error below $10^{-3}$
as expected \cite{stane14b}. This, however, limits the number of bath spins that can be treated in the classical simulation to about $1000$. For significantly larger systems 
the run time becomes too large. In the following sections, we introduce three algorithms that reduce the number of equation of motions even for infinite 
bath sizes  to a manageable size of about $100$ equations.

\section{Expansion of the Overhauser field}
\label{sec.expansion}

In this section, we introduce optimized algorithms to calculate the dynamics of the central spin in classical simulations performed such to be as close
as possible to the quantum mechanical behavior. We start with the hierarchy approach, which uses a hierarchy of Overhauser fields to describe the dynamics. Then, the significantly improved Lanczos approach is developed, 
which uses orthogonal polynomials of the Overhauser fields to overcome problems with the long-time behavior in the hierarchy approach. Finally, we introduce the spectral density approach, which extends the Lanczos approach leading to uniform
convergence of the results in time.

\subsection{Hierarchy approach}
\label{ssec.hierarchy}

In general, for an exact solution of Eqs.\ \eqref{eq:exact_central_spin} and 
\eqref{eq:exact_bath_spin}, one needs to solve $3(N+1)$ coupled differential equations. To reduce the number of equations significantly we
aim at using the Overhauser field as dynamical variable instead of the 
single bath spins. To this end, we introduce the hierarchy of fields
\begin{align}
\vec{B}_n := \sum\limits_{i=1}^{N} J_i^n \vec{S}_i .
\end{align}
Clearly, $\vec{B}_1$ is the original Overhauser field $\vec{B}$.
The dynamics of the hierarchy is given by the straightforward 
equation of motion
\begin{align}
\label{eq:hierarchy}
\frac{\mathrm{d}}{\mathrm{d}t} \vec{B}_n = \vec{S}_0 \times \vec{B}_{n+1} ,
\end{align}
which is exact if one considers the full hierarchy 
$n \in \left\lbrace 1 \dots N \right\rbrace$. A possible truncation, however, cuts 
the hierarchy according to
 $n \in \left\lbrace 1 \dots N_\mathrm{tr} \right\rbrace$ with 
$N_\mathrm{tr} < N$. This neglects the higher Overhauser fields
and treats the last one kept as constant.

While the individual vectors $\vec{S}_i$ are uncorrelated, Gaussian random fields, the Overhauser fields  $\vec{B}_n$ are correlated obeying
\begin{align}
\left\langle B_n^\alpha B_m^\beta \right\rangle = \frac{1}{4} 
\delta_{\alpha\beta} \sum\limits_{i=1}^{N} J_i^{n+m} .
\end{align}
This  symmetric correlation matrix can be mapped to uncorrelated diagonal fields using an orthogonal transformation. In this way, the initial conditions
can be determined from randomly drawn Gaussian variables.

In the numerical simulations, see Sect.\ \ref{sec.results}, it becomes evident
that the hierarchy approach does not converge well. This can be traced back
to the fact that for fixed $\vec{S}_0$ the set of linear 
equations \eqref{eq:exact_bath_spin} can be diagonalized
displaying purely imaginary
eigen values implying oscillatory solutions.
They represent the precession of angular momenta as it has to be.
However, the truncated linear equations \eqref{eq:hierarchy} cannot be diagonalized
and instead of oscillatory solutions we find polynomial behavior,
which approximates the precessions only poorly.

\subsection{Lanczos approach}
\label{ssec.lanczos}

Based on the observation that in the hierarchy approach higher powers of $J_i$
appear in the equations of motion, we develop the Lanczos approach.  We 
introduce uncorrelated fields with polynomials $p_n$ of $J_i$ as prefactors
\begin{align}
\label{eq:def_P_n}
\vec{P}_n := \sum\limits_{i=1}^N p_n(J_i) \vec{S}_i,
\end{align}
where the subscript $n$ denotes the degree of the polynomial.
For simplicity, we assume henceforth that the $J_i$ are normalized 
so that $J_\mathrm{Q}=1$ holds, i.e., they are given relative to $J_\mathrm{Q}$.
In order to have uncorrelated fields, we require that the polynomials are
orthogonal with respect to the scalar product
\begin{align}
\label{eq:scalar}
\left( p_n | p_m \right) := \sum\limits_{i=1}^{N} p_n(J_i) p_m(J_i) = \delta_{nm} .
\end{align}
Then, the correlation matrix is also diagonal
\begin{align}
\left\langle P_n^\alpha P_m^\beta \right\rangle = 
\frac{1}{4} \delta_{nm} \delta_{\alpha\beta},
\end{align}
which is very advantageous, but not yet the key point for introducing
these generalized Overhauser fields, for examples see Eq.\ \eqref{eq:illus} below.

In addition, we construct the polynomials in the usual way by
iterated multiplication of the argument, i.e., by the Lanczos algorithm,
 see Appendix\ \ref{app.lanczos}, implying the recursion
\begin{align}
\label{eq:recursiv}
x p_n(x) = \beta_n p_{n+1}(x) + \alpha_n p_n(x) + \beta_{n-1} p_{n-1}(x)
\end{align}
for $n\in\{1,2,3, \ldots\}$ and starting from $p_0(x):=0$ and $p_1(x):=x$.
The real coefficients $\alpha_n$ and $\beta_n\ge 0$ result from 
the Lanczos iterative determination of the orthogonal polynomials.
Then, the equation of motion for the $\vec{P}_n$ becomes
\begin{subequations}
\label{eq:tridgl}
\begin{align}
\frac{\mathrm{d}}{\mathrm{d}t} \vec{P}_n &= 
\vec{S}_0 \times \sum\limits_{i=1}^N p_n(J_i) J_i \vec{S}_i
\\
\label{eq:lanczos}
 &= \vec{S}_0 \times \left( \beta_n \vec{P}_{n+1} +  \alpha_n \vec{P}_n + \beta_{n-1} \vec{P}_{n-1} \right).
\end{align}
\end{subequations}
The central spin still obeys \eqref{eq:exact_central_spin} and we 
note that the Overhauser field $\vec{B}$ is given by $\vec{P}_1$.

If truncated at finite $N_\mathrm{tr}$, the equations of motion \eqref{eq:tridgl}
are similar to the 
one of the hierarchy approach, but display two crucial advantages.
The first is that the initial values of the polynomial fields $\vec{P}_n$
are \emph{uncorrelated} Gaussian random variables of variance $1/4$.
The second advantage, which is crucial,
is that the set of linear equations \eqref{eq:lanczos}
is diagonalizable for fixed central spin yielding imaginary eigenvalues,
which represent the expected precessions, see next section.

\subsection{Spectral density approach}
\label{ssec.spectral}

The Lanczos approach provides differential equations of the form
\begin{align}
\frac{\mathrm{d}}{\mathrm{d}t} \vec{P}_n &= 
\vec{S}_0 \times \sum\limits_{i=1}^{N_\mathrm{tr}} T_{ni} \vec{P}_i
\end{align}
with the tridiagonal matrix
\begin{align}
\label{eq:tridiag}
\underline{\underline{T}} = 
\begin{pmatrix}
\alpha_1 & \beta_1 & 0 & 0 & \cdots & 0 \\
\beta_1  & \alpha_2 & \beta_2 & 0 & \cdots & 0 \\
0    & \beta_2 & \alpha_3  & \beta_	3 & \ddots & \vdots  \\
0 & \ddots & \beta_3 & \alpha_4 & \ddots & 0 \\
\vdots & \ddots & \ddots & \ddots & \ddots &\beta_{N_\mathrm{tr}-1} \\
0 & 0 & \cdots & 0 &\beta_{N_\mathrm{tr}-1} & \alpha_{N_\mathrm{tr}}
\end{pmatrix} .
\end{align}
The matrix $\underline{\underline{T}}$ is symmetric and real, 
hence it can be diagonalized with real eigen values 
$\varepsilon_\alpha$ and eigen vectors 
$\vec{U}_\alpha \in \mathbb{R}^L$ with 
$\alpha \in \left\lbrace1 \dots N_\mathrm{tr}\right\rbrace$.
Then we can define the diagonal dynamical vectors
\begin{align}
\vec{Q}_\alpha (t) := \sum\limits_{m=1}^{N_\mathrm{tr}} 
(\vec{U}_\alpha)_m \vec{P}_m(t) .
\end{align}
Their equations of motion read
\begin{align}
\frac{\mathrm{d}}{\mathrm{d}t} \vec{Q}_\alpha (t) = \varepsilon_\alpha \vec{S}_0 \times \vec{Q}_\alpha (t),
\end{align}
which is even simpler than before thanks to the diagonalization.
The equation of motion for the central spin is determined by the 
 Overhauser field $\vec{B}$, which equals the first 
polynomial field
\begin{align}
\label{eq:p1}
\vec{P}_1(t) = \sum\limits_{\alpha=1}^{N_\mathrm{tr}} 
(\vec{U}_\alpha)_1 \vec{Q}_\alpha (t) ,	
\end{align}
where we assume that the matrix elements $(\vec{U}_\alpha)_1$
all are non-negative for later use. If not, one can rescale the vectors
$\vec{Q}_\alpha$ appropriately. 
So far this approach is equivalent to the Lanczos approach, except that it is 
expressed in a diagonal basis. The spectral density approach 
goes some steps further, realizing a suitable continuum
limit.

First, we recall from mathematics that 
orthogonal polynomials $q_n(x)$ require a scalar product which
is defined by a weight function $w(x)\ge 0 $ \cite{abram64}
\begin{subequations}
\begin{align}
(f|g) &:= \int w(x) f(x) g(x) \mathrm{d}x
\\
(q_m|q_n) &= \int w(x) q_m(x) q_n(x) \mathrm{d}x
\label{eq:weight}
\\
&= \delta_{mn} .
\label{eq:q-ortho}
\end{align}
\end{subequations}
The only difference between the $p_n$ and the standard definition
is that the $p_n$ start with $p_1=x$ instead of $q_1=1$.
Hence we simply define 
\begin{align}
\label{eq:qdef}
q_n(x) &:= p_n(x)/x.
\end{align}
Furthermore, we recall that the weight function
can be retrieved from the $1,1$ matrix element of the retarded 
resolvent of $\underline{\underline{T}}$ by
\begin{align}
w(x)&=\frac{-1}{\pi} \mathrm{Im} \lim_{\delta\to 0+}
\left(\frac{1}{x+i\delta-\underline{\underline{T}}}\right)_{1,1}.
\end{align}
Expressed in its diagonal basis this equation implies
\begin{align}
\label{eq:wdiag}
w(x) &=\sum_{\alpha=1}^{N_\mathrm{tr}} \left|(\vec{U}_\alpha)_1 \right|^2 
\delta(x-\epsilon_\alpha).
\end{align}

Next, we find the weight function. Since the orthonormality
\eqref{eq:scalar} must be preserved in \eqref{eq:q-ortho}
we deduce
\begin{subequations}
\begin{align}
(p_m|p_n) &= \sum_{i=1}^N p_m(J_i)p_n(J_i)
\\
&= \sum_{i=1}^N J_i^2 q_m(J_i)q_n(J_i)
\label{eq:qscalar}
\\
&= (q_m|q_n)
\end{align}
\end{subequations}
Comparing \eqref{eq:qscalar} with \eqref{eq:weight} reveals
\begin{align}
\label{eq:wdef}
w(x) &:= \sum_{i=1}^N x^2\delta(x-J_i).
\end{align}
Note that the normalization $J_\mathrm{Q}=1$ in \eqref{eq:jnorm} implies
that the integral over $w(x)$ yields unity.

Naturally, the weight function for any finite spin bath consists
of a finite number of $\delta$-peaks. In view of the extremely large number
of bath spins in quantum dots, it makes sense to address a
suitable continuum limit. This can be done by approximating
the discrete sums by integrals. To this end, we start from 
a general distribution of the couplings given by
\begin{align}
J_i =  C f(\gamma i),
\end{align}
where $f(x)$ for $x \in \left[0,x_0\right]$ is a monotonic decreasing function 
starting at $f(0) = 1$ and vanishing at $f(x_0)=0$. 
If $\gamma \ll 1$, we replace 
\begin{subequations}
\begin{align}
\label{eq:weight_discrete}
w(x) &= \sum\limits_{i=1}^{N} x^2\delta(x-C f(\gamma i))
\\
\label{eq:weight_continuum}
&\approx \int_0^{x_0} \frac{x^2}{\gamma}\delta(x-C f(y)) \mathrm{d}y 
\\
&= \left. \frac{x^2}{\gamma C|f'(y)|}\theta(x(C-x)) \right|_{x=Cf(y)} ,
\end{align}
\end{subequations}
where $\theta(x)$ is the Heaviside step function. 
This is the general result. For the exponential parametrization 
\eqref{eq:exponen} the weight function
$w(x)$ is easily computed yielding
\begin{align}
\label{eq:specdens}
w(x) = 	\frac{x}{\gamma}\theta(x(\sqrt{2\gamma}J_\text{Q}-x))  .
\end{align}
This particularly simple spectral density is illustrated in
Fig.\ \ref{fig:specdens}. Further continuous weight functions
are derived in Appendix\ \ref{app.weight} and shown in Fig.\ \ref{fig:weightcompare}.

We point out that the simplicity of the the linear weight
function allows us to provide the tridiagonal coefficients
$\{\alpha_i,\beta_i\}$ analytically as they enter in \eqref{eq:tridiag}.
From the known continued fraction representation for the Jacobi polynomials
\cite{petti85}
we deduce
\begin{subequations}
\begin{align}
\alpha_n &= \frac{4n^2}{4n^2-1}\sqrt{\frac{\gamma}{2}}
\\
\beta_n &= \frac{\sqrt{n(n+1)}}{2n+1}\sqrt{\frac{\gamma}{2}} .
\end{align}
\end{subequations}
This allows us to carry out calculations directly in the
continuum limit based on the Lanczos approach. For large 
enough $N_\text{tr}$ the  results obtained in this way
are numerically exact and will serve as test bed for the 
spectral density (SD) approach. For illustration, we show
the first three generalized Overhauser fields for these
recursion coefficients
\begin{subequations}
\label{eq:illus}
\begin{align}
\vec{P}_1 &= \sqrt{\gamma} \sum_{i=1}^N \overline{J}_i \vec{S}_i
\\
\vec{P}_2 &= \sqrt{\gamma} \sum_{i=1}^N (3\overline{J}_i-\sqrt{8}) \overline{J}_i \vec{S}_i
\\
\vec{P}_3 &= \sqrt{\gamma} \sum_{i=1}^N \sqrt{3} (5\overline{J}_i^2-6\sqrt{2}\overline{J}_i+3) \overline{J}_i \vec{S}_i,
\end{align}
\end{subequations}
where $\overline{J}_i:=J_i/\sqrt{\gamma}$.

\begin{figure}[ht]
\centering
\includegraphics[width=1.0\columnwidth]{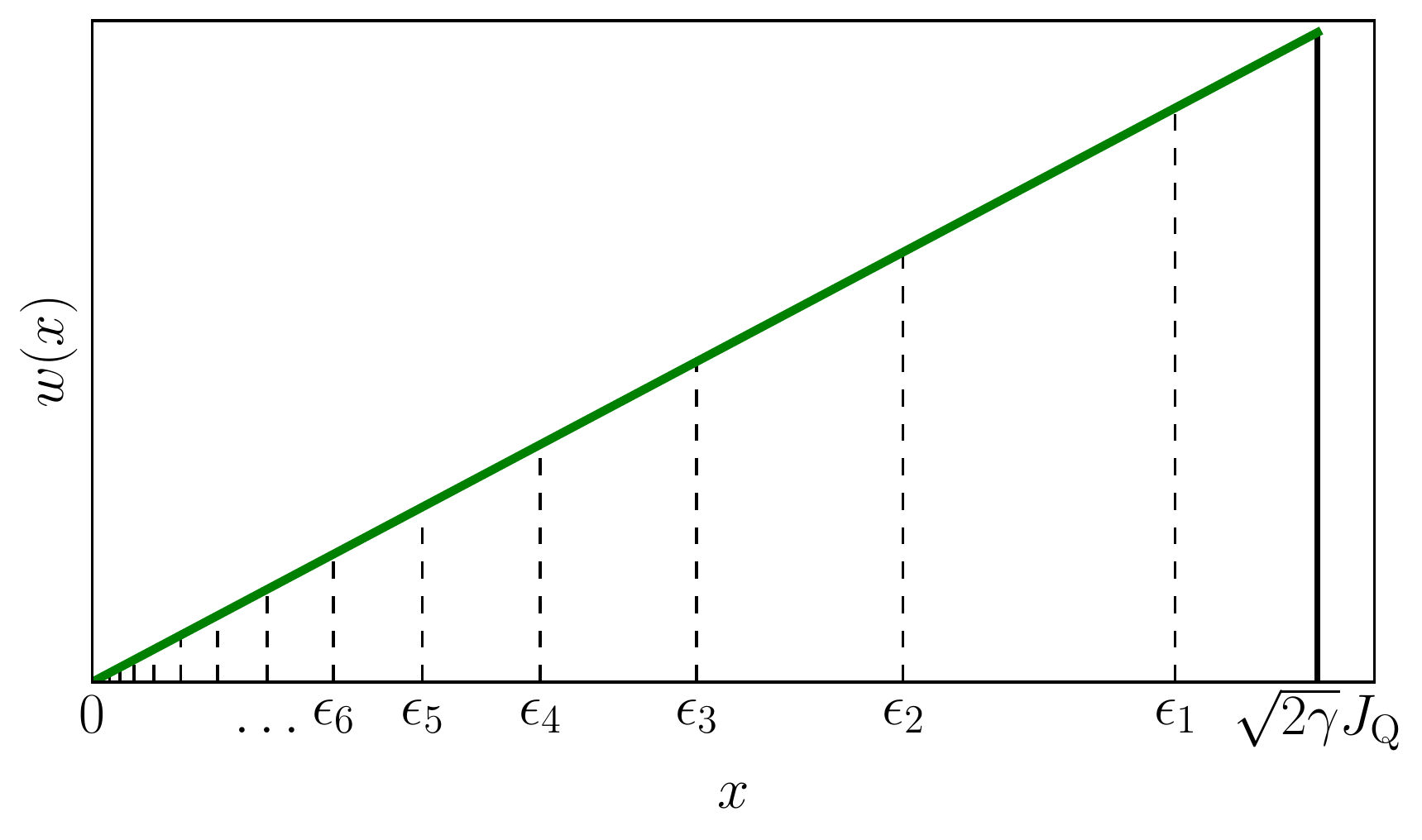}
\caption{
Sketch of the spectral density \eqref{eq:specdens} resulting from
the exponential parametrization \eqref{eq:exponen} in the limit of small $\gamma$.
In addition, the chosen exponential eigen energies are shown as they
result from the procedure explained around Eq.\ \eqref{eq:lambdadef}
for $\lambda=0.8$.}
\label{fig:specdens}
\end{figure}

The SD approach aims at a most efficient representation of the 
continuous spectral density $w(x)$
by a small number of dynamic variables.
Hence, we choose the well-established exponential discretization of
the energies in order to capture the long-time behavior. 
We first define the grid
\begin{align}
\label{eq:epsilondef}
\tilde{\epsilon}_i = \lambda^{i} \left(  
\frac{\epsilon_\mathrm{max}(N_\mathrm{tr}-i)}
{N_\mathrm{tr}} \right),
 \quad i \in  \left\lbrace 0 \dots N_\mathrm{tr} \right\rbrace,
\end{align}
where $N_\mathrm{tr}+1$ is the maximum number of grid points $\tilde{\epsilon}_i$
yielding $N_\mathrm{tr}$ intervals
and $\epsilon_\mathrm{max}$ is the maximum value where $w(x)$ is
finite, i.e., $\epsilon_\mathrm{max}=\sqrt{2\gamma}J_\text{Q}$, 
see Fig.\ \ref{fig:specdens}.
The factor $\lambda < 1$ ensures an exponential zoom towards 
lower frequencies. This factor is chosen according to
\begin{align}
\label{eq:lambdadef}
\lambda = \left( \frac{N_\mathrm{tr}}{\epsilon_\mathrm{max}
t_\mathrm{max}} \right)^{\frac{1}{N_\mathrm{tr}-1}} .
\end{align}
The guiding idea of the above expression is to identify a maximum time 
$t_\mathrm{max}$ up to which we wish to compute the time evolution.
Then, we have to keep the modes with sufficiently low energies
such that they precess at most a fraction of a complete turn, i.e., we set 
$\tilde{\epsilon}_{N_\mathrm{tr}-1}t_\mathrm{max}=1$. This condition
fixes $\lambda$ as given by \eqref{eq:lambdadef}. In rare cases where
\eqref{eq:lambdadef} would yield a value $\lambda>1$ we set $\lambda=1$
refraining from an exponential zoom because the linear discretization
is already sufficient.

Finally, the discretization energies  $\epsilon_i$ are chosen such that they
are the average over $w(x)$ between $\tilde{\epsilon}_{i}$ and 
$\tilde{\epsilon}_{i-1}$
\begin{align}
\epsilon_i &:= 
\left. \int^{\tilde{\epsilon}_{i-1}}_{\tilde\epsilon_i} x \, w(x) \,\mathrm{d} x 
\right/ \int^{\tilde{\epsilon}_{i-1}}_{\tilde\epsilon_i} w(x) \, \mathrm{d} x .
\end{align}
This choice guarantees that the weight and the first moment in each of the intervals
and hence for the total weight function are correctly represented by
the discretization.
The finite set of equations is now given by
\begin{align}
\frac{\mathrm{d}}{\mathrm{d}t} \vec{Q}_i (t) = 
\epsilon_i \vec{S}_0 \times \vec{Q}_i (t), \quad i 
\in  \left\lbrace 1 \dots N_\mathrm{tr}  \right\rbrace.
\end{align}
Equation\ \eqref{eq:exact_central_spin} still holds and thanks
to Eqs.\ \eqref{eq:p1} and \eqref{eq:wdiag} we know that the Overhauser
field $\vec{B}=\vec{P}_1$ can be expressed by
\begin{align}
\vec{P}_1 = \sum\limits_{i=1}^{N_\mathrm{tr}} 
\sqrt{W_i} \vec{Q}_i (t),
\end{align}
where $W_i$ denotes the weight in the interval $i$
\begin{align}
W_i = \int_{\tilde{\epsilon}_{i-1}}^{\tilde{\epsilon}_i} w(x) \, \mathrm{d} x.
\end{align}
Thus, the only free parameter left is the number of intervals $N_\mathrm{tr}$, which must be chosen large enough to reach reliable results. 

We point out that the exponential discretization advocated above
can also be used to efficiently approximate the discrete
weight function defined in \eqref{eq:wdef} for finite baths. 
However, we emphasize that the continuum limit
yields excellent results in view of the large number of bath spins
in quantum dots. Moreover, it has the conceptually advantageous 
features (i) to reduce the number of parameters ($N$ drops out) 
and (ii) to allow for scaling arguments, see below.

\section{Results}
\label{sec.results}

In the previous section, we have proposed three different algorithms, 
which aim at enhancing the performance in the computation of dynamical
correlations in the classical central spin model. This is crucial to reach 
long times for large spin baths. While in the full classical 
simulation the number of differential equations scales proportional to 
$N$, the proposed algorithms 
scale proportional to $N_\mathrm{tr} \ll N$. Here, we analyze how 
$N_\mathrm{tr}$ has to be chosen to obtain 
reliable results. Note that the total number of differential equations 
is given by $3 (N_\mathrm{tr} +1)$ in all three algorithms.
Additionally, we study the dependence of the long-time behavior
on the parameter $\gamma$, which is proportional to the inverse effective
number of bath spins. We retrieve the long-time scale of the slow dynamics
\cite{khaet03,coish04}, which is essentially given by the inverse of the maximum individual coupling $1/J_1$
corresponding to $1/(\sqrt{\gamma}J_\text{Q})$. It is a particular strength of
the advocated real-time approach that the dynamics on this time scale is
accessible.

\subsection{Comparison of all approaches}

All numerical data shown has been averaged over $10^6$ initial configurations
(if not stated otherwise), which are picked at random from Gaussian distributions 
for all spin components. This approximates the quantum dynamics \cite{stane14b}.

\begin{figure}[ht]
\centering
\includegraphics[width=1.0\columnwidth]{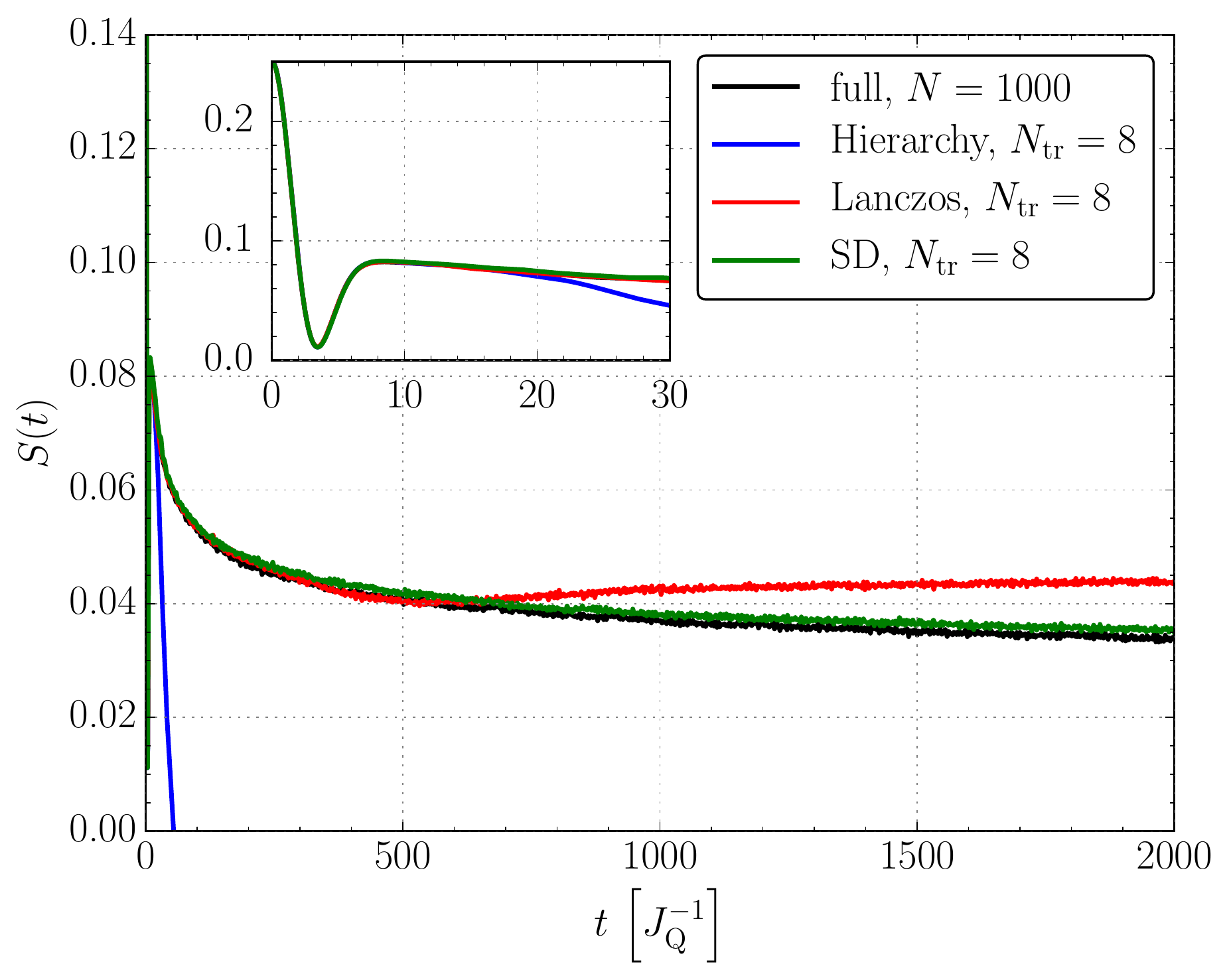}
\caption{Comparison of the hierarchy, the Lanczos, and the spectral density (SD)
approach for fixed truncation parameter $N_\mathrm{tr} = 8$ 
with the solution of the full classical simulation 
for a bath of $N=1000$ spins (except the continuous SD calculation) 
and $\gamma = 0.01$. The inset shows a zoom for short
times.}
\label{fig:1}
\end{figure}

In Fig.\ \ref{fig:1}, we compare the hierarchy, Lanczos and spectral density 
(SD) approach for the fixed truncation parameter $N_\mathrm{tr} = 8$ with the 
numerically exact solution of the full classical simulation for $N=1000$ spins. All 
algorithms capture the short-time dynamics up to $t \approx 15 J_\mathrm{Q}^{-1}$ very well.
However, the hierarchy approach shows a strong deviation already at 
$t \approx 20 J_\mathrm{Q}^{-1}$. Upon increasing $N_\mathrm{tr}$ the
hierarchy results improve, but very slowly. Hence we conclude that this
algorithm is not efficient. We had anticipated this conclusion
already in Sect.\ \ref{ssec.hierarchy} and discussed the reasons for it.
The general mathematical structure of the hierarchy approach is not
appropriate to capture the long-time behavior.

In contrast, both the Lanczos and the SD approach capture the exact solution up 
to remarkably long times in spite of the fairly small truncation parameter. The 
Lanczos method starts to deviate at about $t \approx 700 J_\mathrm{Q}^{-1}$ while 
the SD approach is close to the exact solution for all displayed times.
A deeper understanding of how the results of the Lanczos and the SD approach 
depend on the truncation parameter $N_\mathrm{tr}$ is given in the following 
two subsections.

\subsection{Lanczos approach}

\begin{figure}[ht]
\centering
\includegraphics[width=1.0\columnwidth]{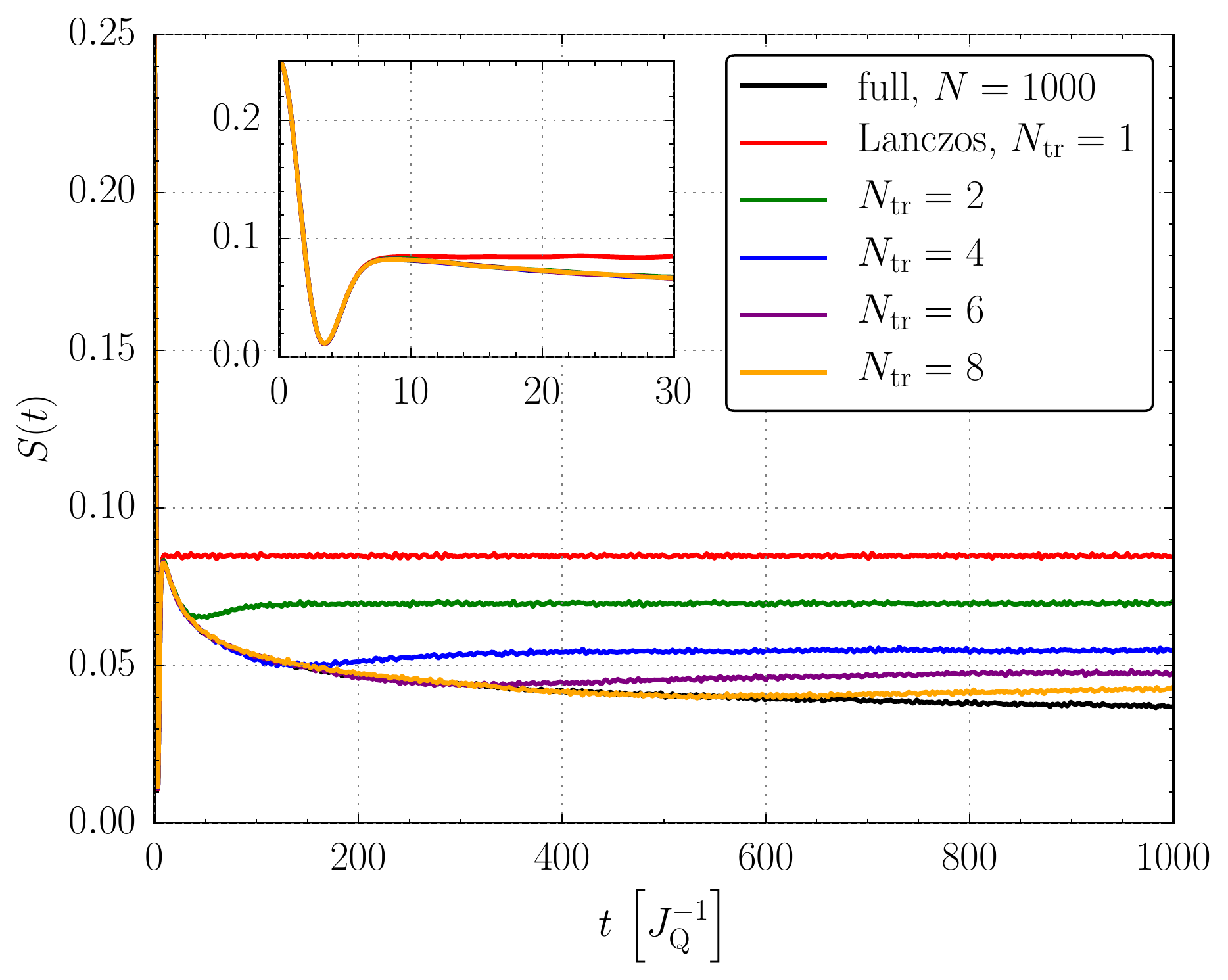}
\caption{Comparison of the Lanczos approach for various truncation 
parameters $N_\mathrm{tr}$ with  the solution of the full classical simulation
for $N=1000$ bath spins and $\gamma = 0.01$.}
\label{fig:2}
\end{figure}

\begin{figure}[ht]
\centering
\includegraphics[width=1.0\columnwidth]{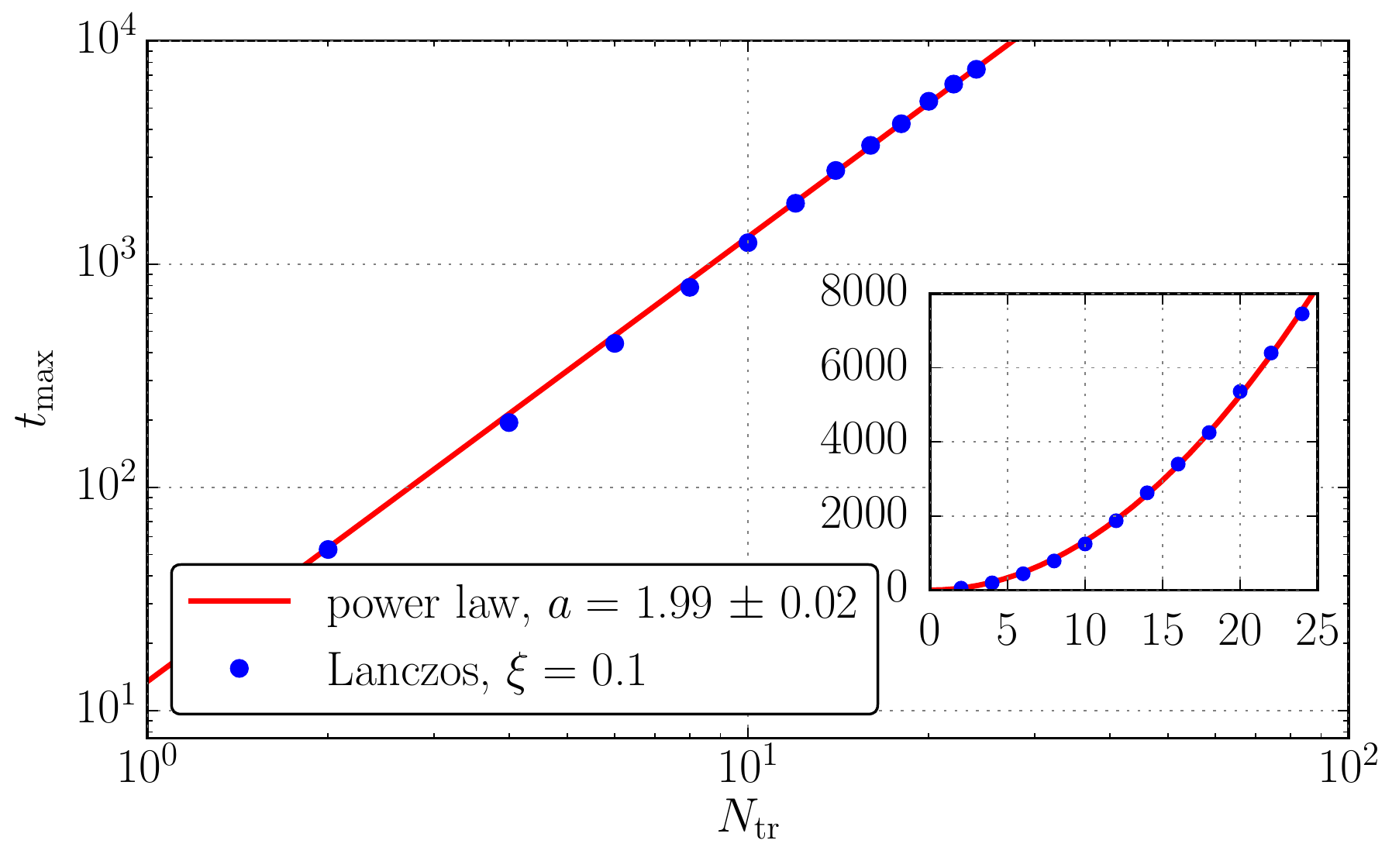}
\caption{Scaling of the time $t_\mathrm{max}$ up to which the Lanczos approach 
is reliable with the truncation parameter $N_\mathrm{tr}$. A power-law fit
$t_\mathrm{max} \propto N_\mathrm{tr}^a$ indicates the exponent $a = 1.99 \pm 0.02$.}
\label{fig:3}
\end{figure}

Figure \ref{fig:2} shows the dependence of the results of the Lanczos approach
on an increasing truncation parameter $N_\mathrm{tr}$. It is obvious that after 
a specific time, the solution starts to deviate from the exact result
and displays a spurious plateau region. Up to the specific time the solution
is very accurate. In order to know beforehand until which time one
may trust the results we introduce the time $t_\mathrm{max}$ at which the 
relative deviation exceeds a certain threshold $\xi$, for instance 
 $\xi = 0.1$. In Fig.\ \ref{fig:3}, we study the dependence of $t_\mathrm{max}$
 on $N_\mathrm{tr}$. A power-law fit $t_\mathrm{max} \propto N_\mathrm{tr}^a$ in
 the double-log plot clearly shows that the scaling is quadratic:
$t_\mathrm{max} \propto N_\mathrm{tr}^2$. 

Therefore, by increasing the truncation parameter $N_\mathrm{tr}$, much longer simulation times can be reached using the Lanczos approach while still having a significant advantage in performance over the full classical simulation.We stress
that increasing $N_\mathrm{tr}$ does not lead to a deterioration of the
description of the short-time dynamics, see inset of Fig.\ \ref{fig:2}.

\subsection{Spectral density approach}
\label{ssec.SD}

\begin{figure}[ht]
\centering
\includegraphics[width=1.0\columnwidth]{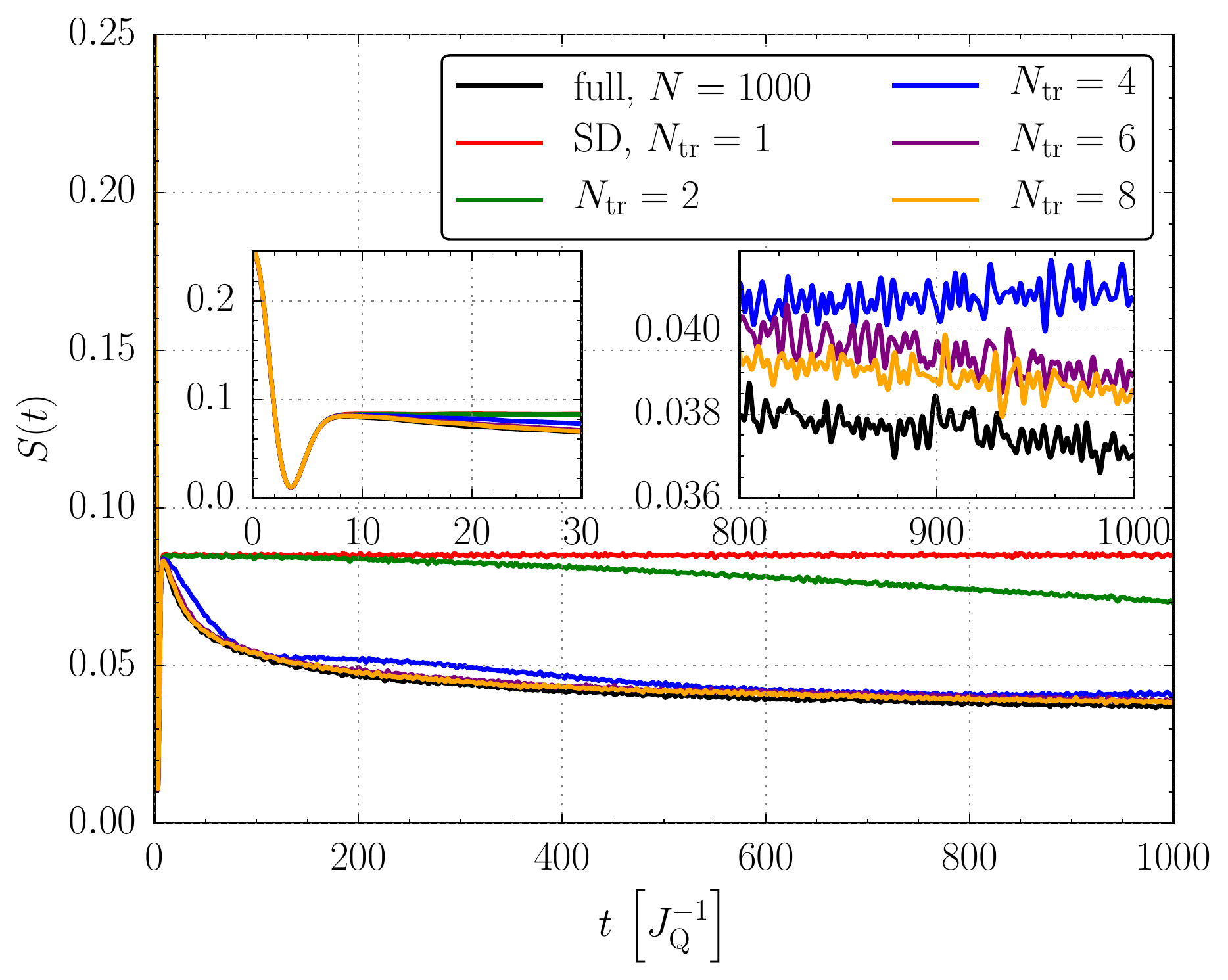}
\caption{Comparison of results from the spectral density approach
 to the full classical simulation for various truncation parameters 
$N_\mathrm{tr}$ at $\gamma = 0.01$. The full simulation is performed for $N=1000$ 
bath spins. The wiggles in the right inset result from the averaging
over $10^6$ initial Gaussian configurations. They scale like the inverse square root
of the number of configurations considered in the average.}
\label{fig:4}
\end{figure}

In contrast to the Lanczos approach, the spectral density (SD) approach 
shows a completely different behavior upon increasing the truncation parameter 
$N_\mathrm{tr}$ as is illustrated in Fig.\ \ref{fig:4}. As for the
Lanczos results the SD results improve upon increasing $N_\mathrm{tr}$.
But the convergence is roughly uniform, i.e., for low values of
$N_\mathrm{tr}$ deviations occur for small and for large times and they
are of about the same magnitude. We consider this to be an
important advantage because we are interested in a faithful 
description for very long times. It is not a particular asset to 
have ultrahigh precision at short times.

The reason for this behavior lies in the particular construction
of the SD algorithm. The energies, which are included in the
description of the bath, are designed to capture all the relevant dynamics
in the time interval under study, see Eqs.\ \eqref{eq:epsilondef}
and \eqref{eq:lambdadef}.

\begin{figure}[ht]
\centering
\includegraphics[width=1.0\columnwidth]{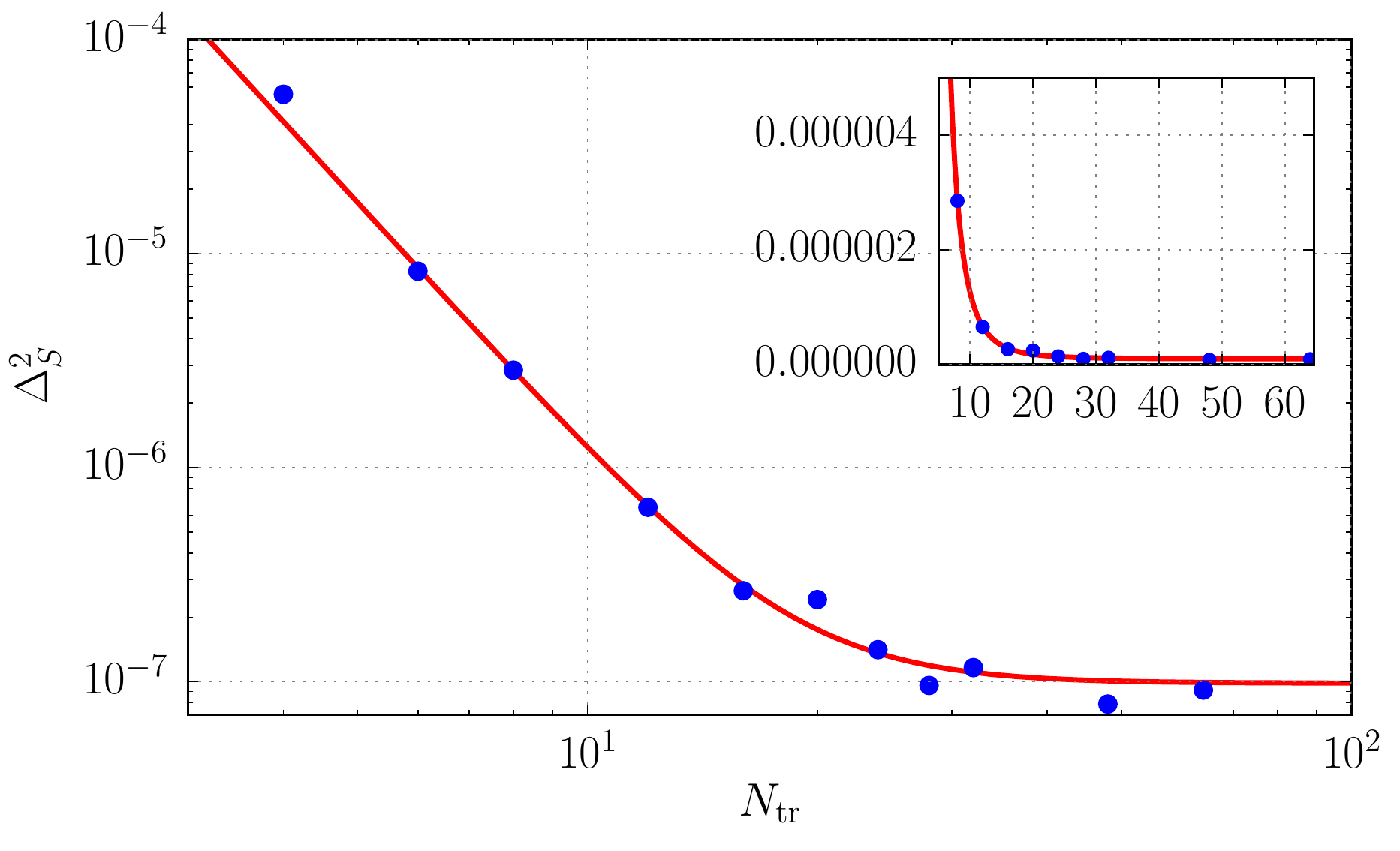}
\caption{Average square difference $\Delta^2_S$ between the
SD result and an accurate Lanczos result in the continuum limit.
The average is computed in the time interval 
$t \in [50, 10000] J_\text{Q}^{-1}$.
The solid line depicts the fit \eqref{eq:fit} which 
is obtained for values from $N_\text{tr}=8$ onwards.
\label{fig:sd_scaling}}
\end{figure}

In order to assess the accuracy of the SD approach quantitatively
we plot in Fig.\ \ref{fig:sd_scaling} the average square difference $\Delta^2_S$
between the SD result and a highly accurate Lanczos calculation in the time interval
under study as function of the truncation parameter $N_\mathrm{tr}$.
Clearly, we see a rapid convergence, which can be fitted by
\begin{equation}
\label{eq:fit}
\Delta^2_S \approx \frac{A}{N_\text{tr}^B} + C
\end{equation}
with $A=0.009 \pm 0.003$, $B= 3.9 \pm 0.1 $, and 
$C = (1.0 \pm 0.1)\cdot 10^{-7}$. The constant offset $C$ occurs naturally
because there remains a statistical error for all $N_\text{tr}$ 
from the average over $10^7$ random Gaussian initial values.
The fit clearly shows that the average convergence is quadratic 
$\Delta_S \propto 1/N_\text{tr}^2$ in 
the inverse number $N_\text{tr}$ of tracked dynamic vectors. 

\begin{figure}[ht]
\centering
\includegraphics[width=1.0\columnwidth]{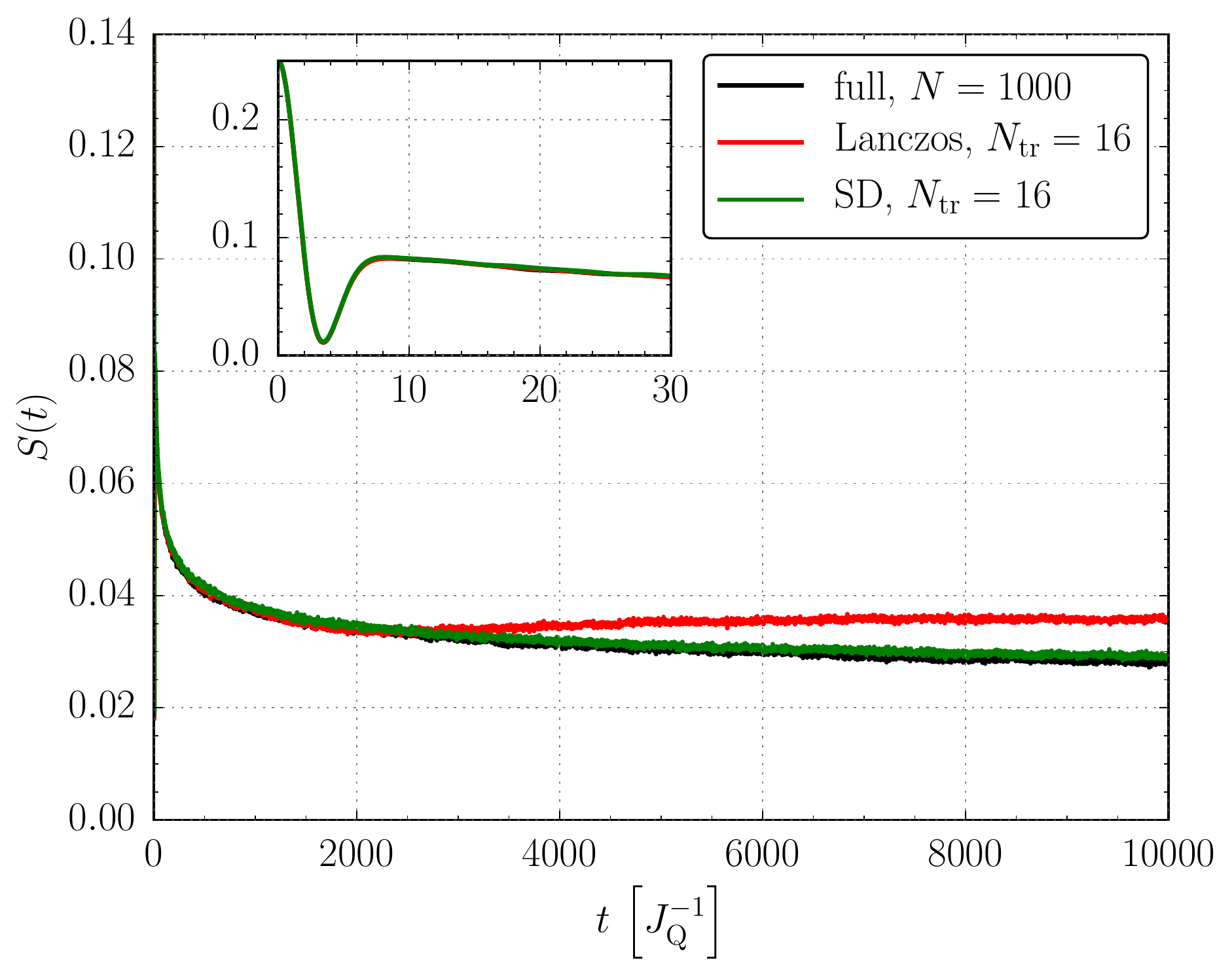}
\caption{Comparison of results from the Lanczos and the spectral density approach
for $N_\mathrm{tr} = 16$ to the full classical simulation for very long times up to $t = 10000 J_\mathrm{Q}^{-1}$ at  $\gamma = 0.01$. The Lanczos and the full calculations are performed for $N=1000$ bath spins.}
\label{fig:5}
\end{figure}

The advantageous feature of the SD approach is summarized in Fig.\ 
\ref{fig:5}, which clearly shows that that the SD approach captures the
dynamics of the central spin model more efficiently than
the Lanczos approach. We stress, however, that the Lanczos approach has the advantage to yield particularly precise data when simulating shorter times.
Both approaches can deal with nominally infinitely large 
spin baths, i.e., for $N=\infty$, while the effective number of bath
spins $N_\mathrm{eff}$ is finite corresponding to a finite parameter 
$\gamma=2/N_\mathrm{eff}$.

\subsection{Long-time behavior}

Above, we have illustrated that the spectral density approach is especially suited 
to describe the dynamics at very long times correctly. Hence, we adopt this
algorithm for the subsequent analysis. We use $N_\mathrm{tr} = 32$ for 
the following calculations. Note that 
the calculations are carried out for infinitely large baths $N=\infty$.

\begin{figure}[ht]
\centering
\includegraphics[width=1.0\columnwidth]{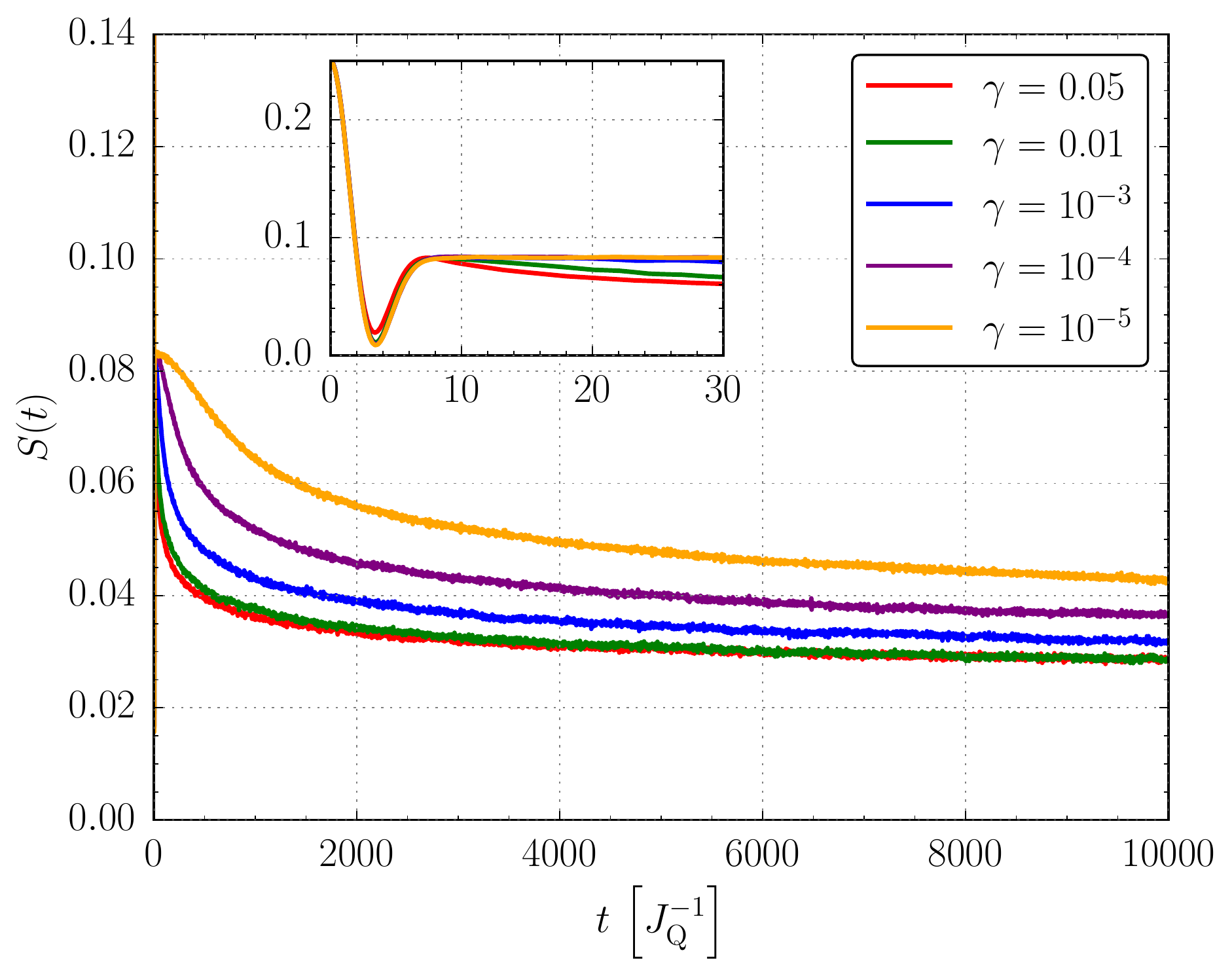}
\caption{Dynamics for various values of $\gamma$ for very long times 
up to $t=10^4 J^{-1}_\text{Q}$ from the
SD approach. The calculations are performed for $N = \infty$ 
bath spins and $N_\mathrm{tr} = 32$.}
\label{fig:6}
\end{figure}

\begin{figure}[ht]
\centering
\includegraphics[width=1.0\columnwidth]{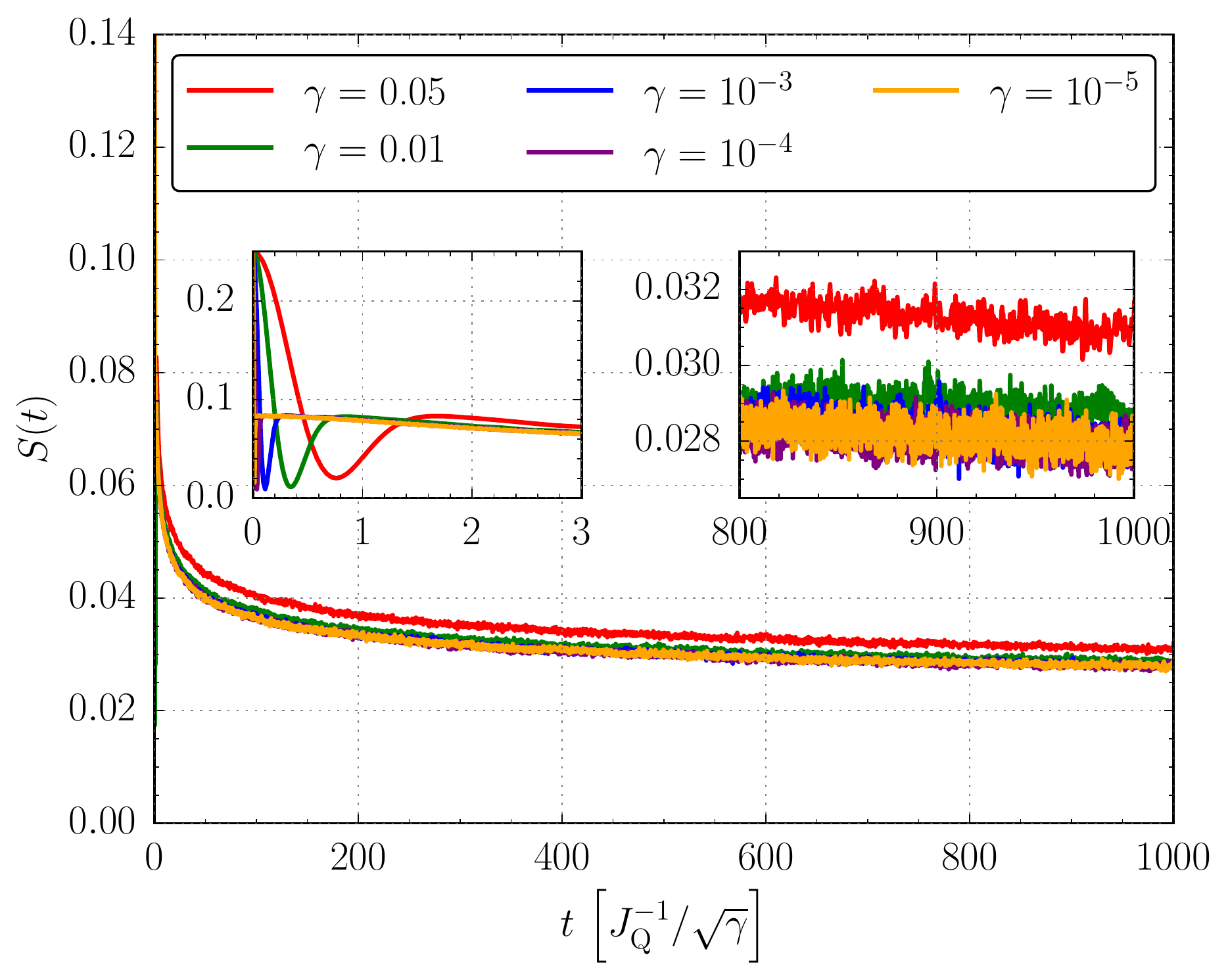}
\caption{Rescaled results from Fig.\ \ref{fig:6} showing an excellent
data collapse as long as the short-time dynamics and the long-time
dynamics are separated clearly, i.e., for $\sqrt{\gamma} \ll 1$.}
\label{fig:7}
\end{figure}

We investigate the influence of the parameter $\gamma$, which represents the inverse number of effectively coupled bath spins.
Figure \ref{fig:6} shows a set of representative results for various 
values of $\gamma$ up to $t=10^4 J_\mathrm{Q}^{-1}$. While the curves are qualitatively very similar, smaller values of $\gamma$ clearly imply a slower long-time
dynamics. We stress that the short-time dynamics, see inset, is not altered
by changing $\gamma$ because it is determined by the energy scale $J_\mathrm{Q}$,
which is the energy scale used throughout this article, i.e., it is set
to unity in the numerics. Hence, all curves coincide in the inset.
Only the $\gamma=0.05$ curve deviates a tiny bit. We attribute this effect 
to the fact that at $\gamma=0.05$ the continuum limit, i.e., the
step from \eqref{eq:weight_discrete} to \eqref{eq:weight_continuum}, does not 
capture the discrete bath perfectly.

Turning back to the long-time dynamics the question arises whether the 
dynamics for different values of $\gamma$ can be mapped on one curve.
This would imply that the information content is essentially the same.
Practically, a good data collapse would help future theoretical simulations
since only moderate values of $\gamma$ need to be analyzed.

Looking at Fig.\ \ref{fig:specdens} and at the analytic result
\eqref{eq:specdens} it is obvious that the maximum energy occurring in the
weight function $w(x)$ sets a second energy scale. This energy scale
is $\sqrt{\gamma}J_\mathrm{Q}$. The first energy scale
is $J_\mathrm{Q}$ as discussed above for the inset of Fig.\ \ref{fig:6}.
Hence, it is natural to assume that the long-time dynamics is 
determined by the second, much smaller energy scale $\sqrt{\gamma}J_\mathrm{Q}$.
To corroborate this hypothesis we plot the data from Fig.\ \ref{fig:6}
with rescaled time argument in Fig.\ \ref{fig:7}. 
Indeed, an impressive data collapse is achieved. In particular for low values
of $\gamma$, the scaling with $\sqrt{\gamma}$ works perfectly.
For larger values of $\gamma$, the two energy scales $J_\mathrm{Q}$
and $\sqrt{\gamma}J_\mathrm{Q}$ are not so clearly separated so that 
the rescaling is not fully quantitative.

Obviously, the short-time dynamics does not match anymore once the rescaling 
with the factor $\sqrt{\gamma}$ has been performed, see inset of
Fig.\ \ref{fig:7}. This is so because the corresponding time scale 
is solely given by $J_\mathrm{Q}^{-1}$.

Another issue is how the correlations in the central spin model
decrease. In the quantum mechanical model we know from rigorous
lower bounds \cite{uhrig14a,seife16} that the correlations
never fade away, but persist even for infinite baths \emph{if}
the couplings are distributed such that their distribution
can be described as probability distribution $p(J)$ with finite moments.
Note that this is not the case for the exponentially parametrized
couplings in \eqref{eq:exponen} and the Gaussian
parametrizations considered in Appendix\ \ref{app.weight} because these
parametrizations imply that there are an infinite number
of very weakly coupled spins. No normalization of a probability
distribution $p(J)$ is possible.

We recall that the lower bounds as discussed in Refs.\ \cite{uhrig14a,seife16}
result from the existence of conserved quantities such as the total
angular momentum and the total energy. These quantities are conserved
also for the classical model and the choices of couplings we 
are considering here. Hence it is not astounding that the correlations
live very long. They are protected by conservation laws and thus
they decrease very slowly as can be seen in Figs.\ \ref{fig:5},  
\ref{fig:6}, and \ref{fig:7}.

\begin{figure}[ht]
\centering
\includegraphics[width=1.0\columnwidth]{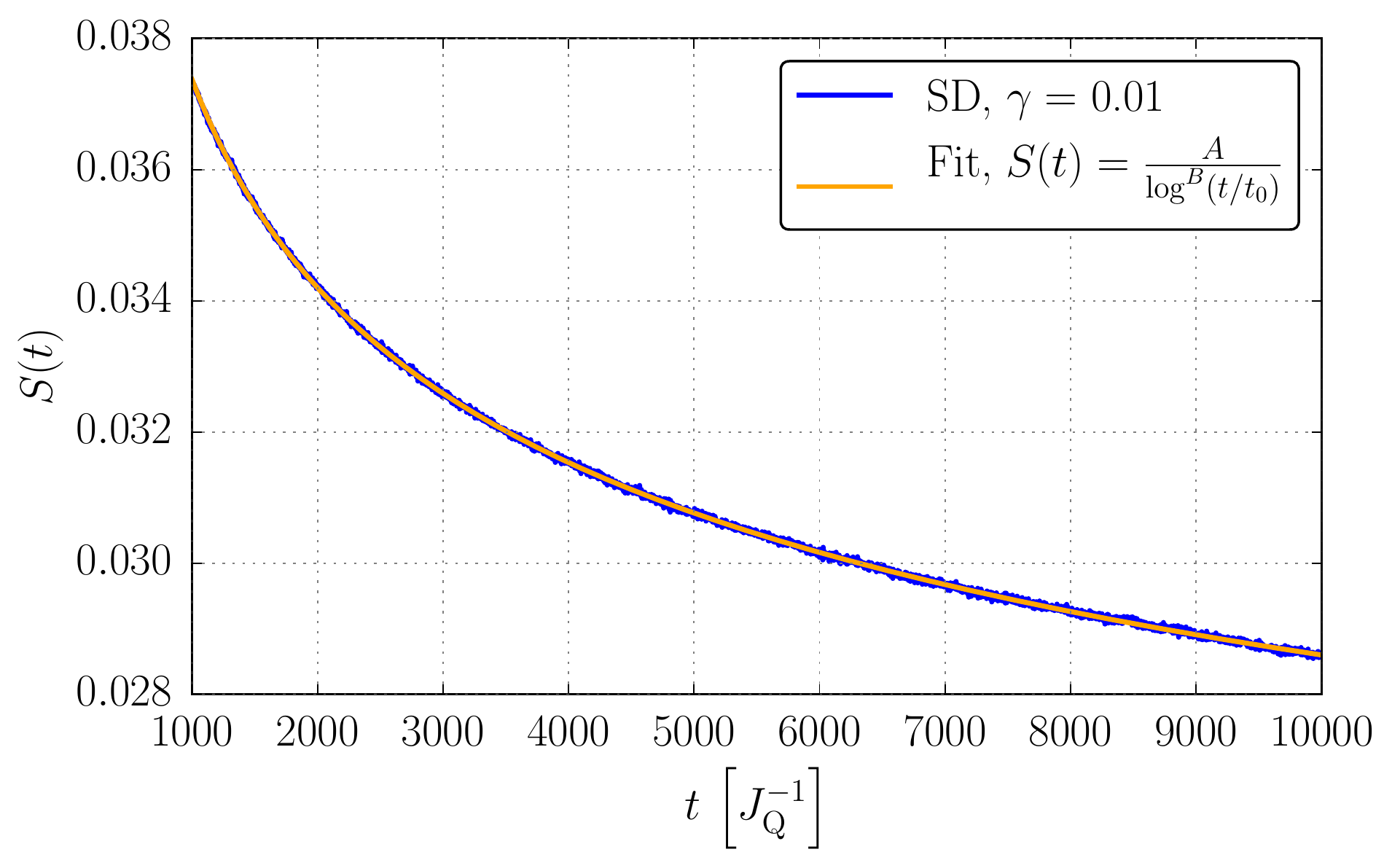}
\caption{Analysis of the long-time behavior of the correlations, computed from $10^8$
initial Gaussian configurations for enhanced accuracy, 
by a logarithmic fit \eqref{eq:logfit}. The parameters are given
in the main text.}
\label{fig:fit}
\end{figure}

The question arises how the slow decay can be described quantitatively.
Chen \textit{et al.}\ proposed a slow logarithmic decay \cite{chen07}. Hence we fit the 
simulated data according to
\begin{align}
\label{eq:logfit}
S(t) = A / \ln^B(t/t_0).
\end{align}
The simulated data and the fit are compared in Fig.\ \ref{fig:fit}.
The fit has been obtained in the interval $t\in[10^3,10^4]J^{-1}_\text{Q}$.
Clearly, it works very well, supporting Chen's suggestion.
The fit parameters are $A = 0.243 \pm 0.001 $, $B = 0.954 \pm 0.002$,
and $t_0 = (0.81 \pm 0.01)/J_\mathrm{Q}$. The fit is
of comparable  quality if we fixed $B=1$. So the existence of a
logarithmic factor is certain, but further details such as the 
precise power, let alone further logarithmic corrections \cite{seife16},
cannot be determined reliably.

There is no need to show and to analyze data for other values of
$\gamma$ due to the above established scaling with $\sqrt{\gamma}$ for
sufficiently small values of $\gamma$.

Finally, we address the influence of varying weight functions.
We do not study wildly different weight functions, but 
stay with plausible choices. The dominant hyperfine coupling
is proportional to the probability of the electron to
be present at the position of the nuclear spins in the quantum dot
\cite{merku02,schli03}. Assuming to first approximation
a parabolic trapping potential as it results from any Taylor
expansion, Gaussian wave functions are the most plausible 
assumption. In Appendix\ \ref{app.weight}, we compute
the three corresponding weight functions $w_d(x)$
in dimension $d=1, 2$, and $3$. The two-dimensional (2D) case is covered
by the linear weight function, which we have used so far.

\begin{figure}[ht]
\centering
\includegraphics[width=1.0\columnwidth]{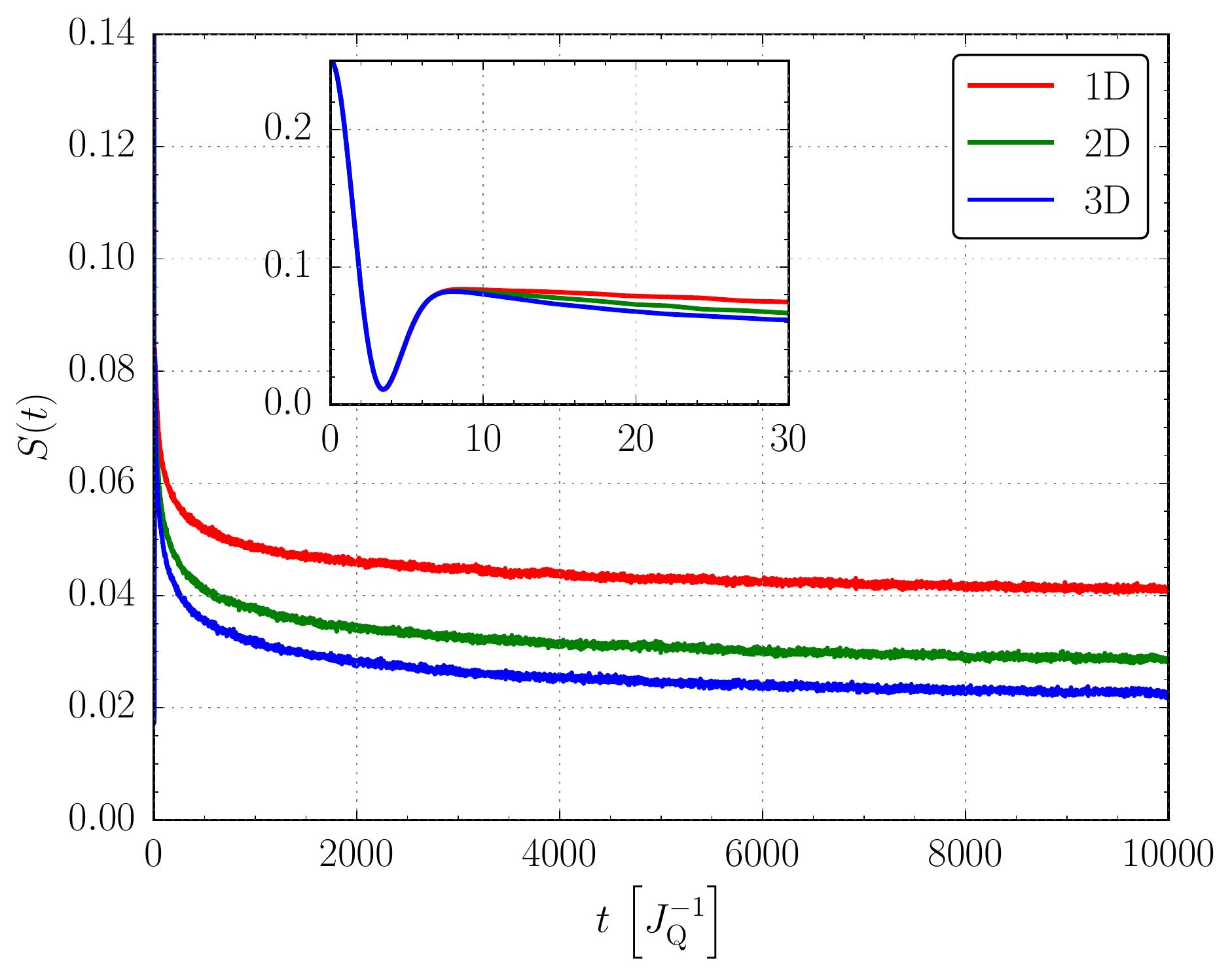}
\caption{Comparison of the dynamics of the spin-spin correlation
in the central spin model for three generic weight functions $w_d(x)$
relevant in dimension $d=1$ (1D), $d=2$ (2D), and $d=3$ (3D) for $\gamma=0.01$. The weight functions
are given in Eqs.\ \eqref{eq:w1}, \eqref{eq:w2}, and \eqref{eq:w3}.
Note that the 2D Gaussian case is identical to the linear
weight function $w(x)$ in \eqref{eq:specdens}.}
\label{fig:weight}
\end{figure}

In Fig.\ \ref{fig:weight},
we compare the resulting dynamic correlation for the
same value of $\gamma=0.01$, which implies the same
number of effectively coupled bath spins.
The results indicate that the influence of the dimensionality
is only moderately important. The main feature of a very slowly,
logarithmically decaying correlation is found in 
all dimensions. The same is true for the scaling 
$\sqrt{\gamma}\propto 1/\sqrt{N_\mathrm{eff}}$.

\section{Conclusions}
\label{sec.conclusions}

We considered the central spin model as relevant description of
two-level systems coupled to large spin baths. 
While the quantum model is the ultimate aim in order to
describe the experimental results, it has been shown that
classical simulations averaged appropriately over  
Gaussian distributed initial conditions provide very good
approximations \cite{alhas06,chen07,stane14b}. Thus 
our study aimed at establishing efficient approaches to
deal with the averaged classical central spin model.
Two demanding challenges had to be met: very large numbers of
bath spins and very long times.

Instead of addressing single spins we introduced generalized higher 
Overhauser fields, which promise to yield a much more
efficient approach. A first attempt, the hierarchy approach,
failed due to an inappropriate mathematical structure.
However, the Lanczos and the spectral density (SD) approach turned out to be
extremely powerful because they require to track  only 10-100 
vectors. The number of these vectors, denoted $N_\mathrm{tr}$, is the control
parameter of the accuracy of the approaches.

The Lanczos approach displays a non-uniform
convergence being excellent up to a certain threshold in time 
$t_\mathrm{max}$, which can be pushed higher and higher by increasing 
$N_\mathrm{tr}$. The scaling is $t_\mathrm{max}\propto N_\mathrm{tr}^2$.
The Lanczos approach is particularly well suited if high-precision data
is required for not too long times.

The SD approach is adjusted to a pre-set time interval.
Within this interval it displays a uniform quadratic convergence
at very moderate computational cost. Moreover, it is based
on the appealing concept of a continuum limit, which 
amounts to setting the \emph{total} number of bath spins
to infinity while the number $N_\mathrm{eff}$ of 
sizeably coupled bath spins within the localization volume
of the electronic central spin is kept as relevant parameter.
In order to establish an appropriate continuum limit,
we introduced weight functions and determined them in
the generic cases.

Employing the powerful SD approach we identified the
energy scale, which is responsible for the long-time behavior.
This low-energy scale is given by $\propto J_\mathrm{Q}/\sqrt{N_\mathrm{eff}}$
where $J_\mathrm{Q}$ is the root of the square sum of all 
couplings. The energy $J_\mathrm{Q}$ is known to dominate the short-time
behavior \cite{merku02,stane13}.
The low-energy scale $J_\mathrm{Q}/\sqrt{N_\mathrm{eff}}$ has appeared
in previous investigations \cite{khaet03,coish04}. 
However, we emphasize that the above introduced algorithm
can produce reliable real-time data up to these
very long time scales for infinite baths with very large
effective number of nuclear spins. This allowed us to 
show by explicit and systematically controlled
calculation that the rescaling of the long-time tails 
of the spin-spin correlation with the low-energy scale 
achieves a convincing data collapse, see Fig.\ \ref{fig:7}.

Physically, the low-energy scale  $J_\mathrm{Q}/\sqrt{N_\mathrm{eff}}$
is obviously a representative value of the individual couplings
of the bath spins. This observation appears to be highly plausible
because the individual bath spin $i$ can react to the behavior of the
central spin only by the rate $J_i/\hbar$. As long as the bath
itself remains static the spin-spin correlation of the central 
spin does not decay but remains constant at one third of its
initial value \cite{merku02,stane13}. Hence, further decay will be slow
and can happen only at the rate at which the bath spin precess.

Finally, we studied the influence of the dimensionality
of the electronic wave function by computing the 
different weight functions $w_d(x)$. The resulting dynamics,
however, indicates only a moderate dependence on the 
dimensionality. This observation also implies that the
details of the couplings in a quantum dot do not matter
much. The key parameters are the high-energy and the low-energy
scale dominating the short-time and the long-time behavior, respectively.

The established approaches and the above observations provide
a reliable algorithmic and conceptual foundation for many 
further investigations. The approach is straightforwardly extended
to finite external magnetic fields acting on the central spin or on the bath spins.
In particular, studies of pulsed quantum dots in external magnetic fields are called for 
\cite{greil06b,greil07a,petro12,econo14,beuge16,beuge17}.

\begin{acknowledgments}
We thank Frithjof B. Anders, Wouter Beugeling, and
Natalie J\"aschke for many helpful discussions. This study has been
supported financially by the Deutsche Forschungsgemeinschaft and the Russian Foundation for Basic Research in International Collaborative Research Centre TRR 160.
\end{acknowledgments}

%\bibliographystyle{apsrev}
%\bibliography{../../bibinput/liter10}

%

\appendix

\section{Lanczos approach}
\label{app.lanczos}

The starting point is $p_1(x) = x$, which is a bit unusual
compared to standard orthogonal polynomials which start at 
$p_1=1$. For the iteration we assume that the recursion
\begin{align}
\label{eq:recursiv2}
x p_m(x) = \beta_m p_{m+1}(x) + \alpha_m p_m(x) + \beta_{m-1} p_{m-1}(x)
\end{align}
for orthonormalized $p_m$ holds up to $m=n-1$. The next step of the induction
iterates $\tilde{p}_{n+1} := x p_n$ where the tilde indicates that
this polynomial is not yet the next orthonormalized polynomial.
The overlaps with already defined polynomials read
\begin{align}
\left( \tilde{p}_{n+1} | \tilde{p}_{n-1} \right) = 
\left( p_{n} x p_{n-1} \right) = \beta_{n-1}.
\end{align}
We compute and define
\begin{subequations}
\begin{align}
\alpha_n &:=\left( \tilde{p}_{n+1} | {p}_{n} \right) 
\\
\beta_n &:=\sqrt{\left| \tilde{p}_{n+1}	- \alpha_n p_n - 
\beta_{n-1} p_{n-1} \right|^2}
\\
p_{n+1}(x) &:=\frac{1}{\beta_n} 
\left( \tilde{p}_{n+1}	- \alpha_n p_n - \beta_{n-1} p_{n-1} \right) . 
\end{align}
\end{subequations}
A straightforward calculation confirms that $p_{n+1}$ defined
in this way obeys \eqref{eq:recursiv2} for $m=n$ and is
orthonormalized with respect to all previously defined polynomials.

We stress that the above construction does not require
that the spin bath is finite. As long as the scalar product
in \eqref{eq:scalar} is well defined, i.e., converges, the
Lanczos approach works.

\section{Weight Functions}
\label{app.weight}

In the main text we established the linear weight function
\eqref{eq:specdens} implied by the exponentially parametrized
couplings \eqref{eq:exponen}. Here we supplement this finding
by three other generic Gaussian parametrizations of the couplings.

\subsection{One-dimensional Gaussian parametrization}

\begin{figure}[ht]
\centering
\includegraphics[width=1.0\columnwidth]{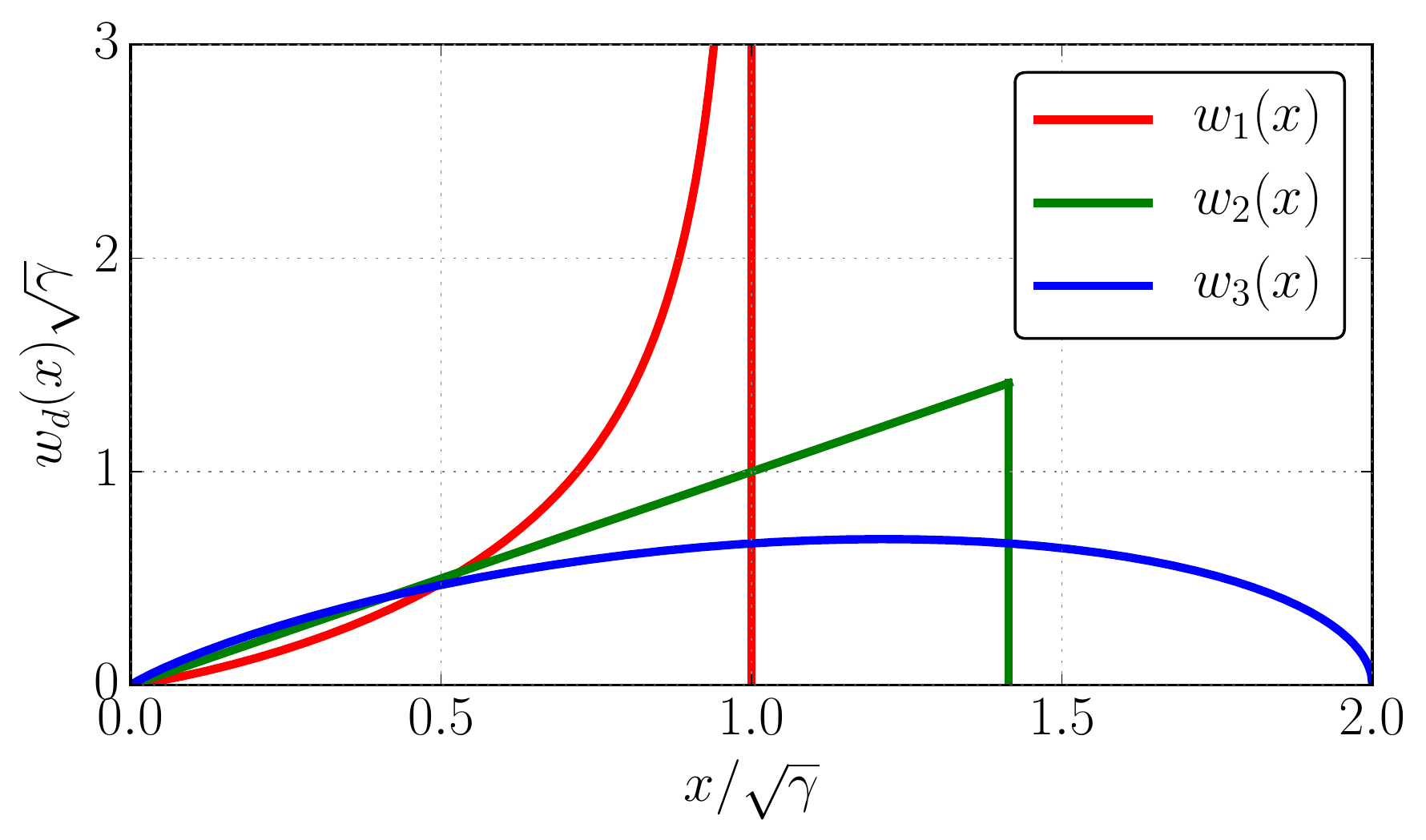}
\caption{Comparison of three weight functions $w_d(x)$ 
resulting from Gaussian parametrized couplings in $d$ dimensions.
The linear weight function of the two-dimensional Gaussian equals
the one of a one-dimensional exponential as discussed in the main
text. This fact underlines the relevance of the linear weight
function.}
\label{fig:weightcompare}
\end{figure}

We consider 
\begin{align}
J_i = C\exp(-\alpha^2 i^2)
\end{align}
with $i\in\{1,2,3,\ldots\}$. For small values of $\alpha$ 
it is justified to approximate the 
sums over all couplings by integrals.
For the energy scale $J_\mathrm{Q}^2$ we obtain
\begin{subequations}
\begin{align}
J_\mathrm{Q}^2 &= \sum_i J_i^2
\\
&= \frac{C^2}{\alpha} \int_0^\infty \exp(-2y^2) \mathrm{d} y
\\
&= \frac{C^2\sqrt{2\pi}}{4\alpha}.
\end{align}
\end{subequations}
Analogously, one obtains
\begin{subequations}
\begin{align}
J_\mathrm{S} &= \sum_i J_i
\\
&= \frac{C}{\alpha} \int_0^\infty \exp(-y^2) \mathrm{d} y
\\
&= \frac{C\sqrt{\pi}}{2\alpha}
\end{align}
\end{subequations}
so that the number of effective bath spins 
$N_\mathrm{eff} = J_\mathrm{S}^2/J_\mathrm{Q}^2$
is given by
\begin{align}
N_\mathrm{eff} &= \sqrt{\frac{\pi}{2}}\frac{1}{\alpha}.
\end{align}
Hence, setting $\alpha=\sqrt{\frac{\pi}{8}} \gamma$ implies
$\gamma=2/N_\mathrm{eff}$ as before. We opt for this choice
of $\alpha$ for better comparability. The energy constant
$C$ results to be
\begin{align}
C &= \sqrt{\gamma}J_\mathrm{Q}
\end{align}
and the weight function to be
\begin{subequations}
\begin{align}
w_1(x) &= \frac{x^2}{\alpha} \int_0^\infty
\delta(x-C\exp(-y^2)) \mathrm{d}y
\\
&= \frac{x\theta(x(C-x))}{\gamma\sqrt{\pi\ln(C/x)/2}}.
\label{eq:w1}
\end{align}
\end{subequations}
It is compared in Fig.\ \ref{fig:weightcompare} to other
 weight functions with the same value of $\gamma$, i.e.,
the same number of effectively coupled spins.

\subsection{Two-dimensional Gaussian parametrization}

We consider 
\begin{align}
J_r = C\exp(-\alpha^2 r^2)
\end{align}
where $r$ is a two-dimensional vector $r\in\mathds{Z}^2$.
For small values of $\alpha$ it is justified to approximate the 
sums over all couplings by integrals.
For the energy scale $J_\mathrm{Q}^2$ we obtain
\begin{subequations}
\begin{align}
J_\mathrm{Q}^2 &= \sum_r J_r^2
\\
&= \frac{2\pi C^2}{\alpha} \int_0^\infty y\exp(-2y^2) \mathrm{d} y
\\
&= \frac{\pi C^2}{2\alpha}.
\end{align}
\end{subequations}
Analogously, one obtains
\begin{subequations}
\begin{align}
J_\mathrm{S} &= \sum_r J_r
\\
&= \frac{2\pi C}{\alpha} \int_0^\infty y\exp(-y^2) \mathrm{d} y
\\
&= \frac{\pi C}{\alpha}
\end{align}
\end{subequations}
so that the number of effective bath spins 
$N_\mathrm{eff} = J_\mathrm{S}^2/J_\mathrm{Q}^2$
is given by
\begin{align}
N_\mathrm{eff} &= \frac{2\pi}{\alpha}.
\end{align}
Hence, setting $\alpha=\pi \gamma$ implies
$\gamma=2/N_\mathrm{eff}$ as before for better comparability. 
The energy constant $C$ results to be
\begin{align}
C &= \sqrt{2\gamma}J_\mathrm{Q}
\end{align}
and the weight function reads
\begin{subequations}
\begin{align}
w_2(x) &= \frac{2\pi x^2}{\alpha} \int_0^\infty
y\delta(x-C\exp(-y^2)) \mathrm{d}y
\\
&= \frac{x\theta(x(C-x))}{\gamma}.
\label{eq:w2}
\end{align}
\end{subequations}
We note that the Gaussian couplings in two dimensions yield
precisely the same weight function as the exponential couplings
\eqref{eq:exponen} in one dimension, see also Fig.\ \ref{fig:weightcompare}.

\subsection{Three-dimensional Gaussian parametrization}

We consider 
\begin{align}
J_r = C\exp(-\alpha^2 r^2)
\end{align}
where $r$ is a three-dimensional vector $r\in\mathds{Z}^3$.
For small values of $\alpha$ it is justified to approximate the 
sums over all couplings by integrals.
For the energy scale $J_\mathrm{Q}^2$ we obtain
\begin{subequations}
\begin{align}
J_\mathrm{Q}^2 &= \sum_r J_r^2
\\
&= \frac{4\pi C^2}{\alpha} \int_0^\infty y^2\exp(-2y^2) \mathrm{d} y
\\
&= \frac{C^2}{\alpha} \left(\frac{\pi}{2}\right)^{3/2}.
\end{align}
\end{subequations}
Analogously, one obtains
\begin{subequations}
\begin{align}
J_\mathrm{S} &= \sum_r J_r
\\
&= \frac{4\pi C}{\alpha} \int_0^\infty y^2\exp(-y^2) \mathrm{d} y
\\
&= \frac{\pi^{3/2} C}{\alpha}
\end{align}
\end{subequations}
so that the number of effective bath spins 
$N_\mathrm{eff} = J_\mathrm{S}^2/J_\mathrm{Q}^2$
is given by
\begin{align}
N_\mathrm{eff} &= \frac{(2\pi)^{3/2}}{\alpha}.
\end{align}
Hence, setting $\alpha= (2\pi)^{3/2} \gamma/2$ implies
$\gamma=2/N_\mathrm{eff}$ as before for better comparability. 
The energy constant $C$ results to be
\begin{align}
C &= 2\sqrt{\gamma}J_\mathrm{Q}
\end{align}
and the weight function reads
\begin{subequations}
\begin{align}
w_3(x) &= \frac{4\pi x^2}{\alpha} \int_0^\infty
y^2\delta(x-C\exp(-y^2)) \mathrm{d}y
\\
&= \sqrt{\frac{2}{\pi}} \frac{x}{\gamma} \theta(x(C-x))\sqrt{\ln(C/x)}.
\label{eq:w3}
\end{align}
\end{subequations}
This  weight function is compared to the other
generic ones in  Fig.\ \ref{fig:weightcompare}.
We note that the differences are not very large
since they result from square roots of logarithmic factors
only.

\end{document}